# High-Dimensional Gaussian Copula Regression: Adaptive Estimation and Statistical Inference[1]

T. Tony Cai and Linjun Zhang
Department of Statistics
The Wharton School
University of Pennsylvania

December 4, 2015

**Abstract**

We develop adaptive estimation and inference methods for high-dimensional Gaussian copula regression that achieve the same performance without the knowledge of the marginal transformations as that for high-dimensional linear regression. Using a Kendall's tau based covariance matrix estimator, an $\ell_1$ regularized estimator is proposed and a corresponding de-biased estimator is developed for the construction of the confidence intervals and hypothesis tests. Theoretical properties of the procedures are studied and the proposed estimation and inference methods are shown to be adaptive to the unknown monotone marginal transformations. Prediction of the response for a given value of the covariates is also considered. The procedures are easy to implement and perform well numerically. The methods are also applied to analyze the Communities and Crime Unnormalized Data from the UCI Machine Learning Repository.

[1]Department of Statistics, The Wharton School, University of Pennsylvania. The research of Tony Cai was supported in part by NSF Grants DMS-1208982 and DMS-1403708, and NIH Grant R01 CA127334.

# 1 Introduction

Finding the relationship between a response and a set of covariates is a ubiquitous problem in scientific studies. Linear regression analysis, which occupies a central position in statistics, is arguably the most commonly used method. It has been well studied in both the conventional low-dimensional and contemporary high-dimensional settings. However, the assumption of linear relationship between the predictors and the response is often too restrictive and unrealistic. Data transformations, such as the Box-Cox transformation, Fisher's $z$ transformation, and variance stabilization transformation, have been frequently used to improve the linear fit and to correct violations of model assumptions such as constant error variance. These transformations are often required to be prespecified before applying the linear regression analysis. See, for example, Carroll and Rupert [7] for detailed discussions on transformations.

For a response $Y$ and predictors $X_1, ..., X_p$, the following functional form of the relationship has been widely used in a range of applications,

$$f_{\lambda_0}(Y) = \boldsymbol{\beta}_0 + \sum_{j=1}^{p} \boldsymbol{\beta}_f f_{\lambda_j}(X_j) + \epsilon, \tag{1}$$

where $f_{\lambda_j}(\cdot)$ are univariate functions and $\lambda_j$ is the parameter associated with $f_{\lambda_j}$. Examples of this model include the additive regression model, single index model, copula regression model, and semiparametric proportional hazards models [9, 20, 21, 23, 26, 30, 33, 40–42]. For applications in econometrics, computational biology, criminology, and natural language processing, see for example [14, 19, 22, 29, 38]. In particular, [42] and [40] established the convergence rates for the minimax estimation risk under the high-dimensional additive regression model and single index model respectively. [27] proposes a plug-in approach for estimating a regression function based on copulas, and presents the asymptotic normality of the estimator. Their model and analysis are restricted to the low-dimensional setting and not well adapted to the high-dimensional case. For data transformations, it is natural to consider the transformations that are continuous and one to one on an interval. Indeed, the functions satisfying these two conditions must be strictly monotonic [36].

In the present paper, we consider adaptive estimation and statistical inference for high-dimensional sparse Gaussian copula regression. The model can be formulated as follows. Suppose we have an independent and identically distributed random sample $\boldsymbol{Z}_1 = (Y_1, \boldsymbol{X}_1)$, ..., $\boldsymbol{Z}_n = (Y_n, \boldsymbol{X}_n) \in \mathbb{R}^{p+1}$ where $Y_i \in \mathbb{R}$ are the responses and $\boldsymbol{X}_i \in \mathbb{R}^p$ are the covariates. Set $d = p + 1$. We say $(Y_i, \boldsymbol{X}_i)$ satisfies a Gaussian copula regression model, if there exists a set of strictly increasing functions $\boldsymbol{f} = \{f_0, f_1, ..., f_p\}$ such that the marginally transformed random vectors $\tilde{\boldsymbol{Z}}_i = (\tilde{Y}_i, \tilde{\boldsymbol{X}}_i) \overset{def}{=} (f_0(Y_i), f_1(X_{i1}), ..., f_p(X_{ip}))$ satisfy $\tilde{\boldsymbol{Z}}_i \overset{i.i.d}{\sim} N_d(0, \Sigma)$ for some positive-definite covariance matrix $\Sigma \in \mathbb{R}^{d\times d}$ with $\text{diag}(\Sigma) = \mathbf{1}$. The condition $\text{diag}(\Sigma) = \mathbf{1}$ is for identifiability because the scaling and shifting are absorbed in the marginal transformations. Note that under the Gaussian copula regression model, one has the following linear relationship for the transformed data:

$$\tilde{Y}_i = \tilde{\boldsymbol{X}}_i^{\top} \boldsymbol{\beta} + \epsilon_i, \quad i = 1, 2, ..., n, \tag{2}$$

where $\boldsymbol{\beta} \in \mathbb{R}^p$ and $\epsilon_i$ are i.i.d zero-mean Gaussian variables. Writing in terms of the covariances, one has $\boldsymbol{\beta} = \Sigma_{\tilde{X}\tilde{X}}^{-1}\Sigma_{\tilde{X}\tilde{Y}}$ and $\epsilon_i \overset{i.i.d}{\sim} N(0, 1 - \Sigma_{\tilde{Y}\tilde{X}}\Sigma_{\tilde{X}\tilde{X}}^{-1}\Sigma_{\tilde{X}\tilde{Y}})$, where $\Sigma_{\tilde{X}\tilde{X}} = \text{Cov}(\tilde{\boldsymbol{X}}_1, \tilde{\boldsymbol{X}}_1)$ and $\Sigma_{\tilde{X}\tilde{Y}} = \text{Cov}(\tilde{\boldsymbol{X}}_1, \tilde{Y}_1)$. We focus on the high-dimensional setting where $p$ is comparable to or much larger than $n$ and $\boldsymbol{\beta}$ is sparse. The fundamental difference between the Gaussian copula regression model and the conventional linear regression model (2) is that one observes $\{Y_1, \boldsymbol{X}_1), ..., (Y_n, \boldsymbol{X}_n)\}$, not $\{(\tilde{Y}_1, \tilde{\boldsymbol{X}}_1), , , , , (\tilde{Y}_n, \tilde{\boldsymbol{X}}_n)\}$ as the transformations $f_i$ are unknown.

The goal of the present paper is to develop adaptive estimation and inference methods that achieve the same performance in terms of the convergence rates without the knowledge of the marginal transformations as that for the high-dimensional linear regression. The rank-based Kendall's tau is used to extract the covariance information on the transformed data that does not require estimation of the transformations. Based on the covariance matrix estimator, an $\ell_1$ regularized estimator is proposed to estimate $\boldsymbol{\beta}$ and a corresponding de-biased estimator is developed for the construction of the confidence intervals and hypothesis tests. In addition, prediction of the response for a given value of the covariates is also considered. Theoretical properties of the procedures for estimation, prediction, and statistical inference are studied. The proposed estimator is shown to be rate-optimal under regularity conditions. The proposed estimation and inference methods share similar properties as those optimal procedures for the high-dimensional linear regression. They are more flexible in the sense that they are adaptive to unknown monotone marginal transformations. For example, it is of practical interest to test whether a given covariate $X_i$ is related to the response $Y$. The proposed testing procedure enables one to test this hypothesis without the need of knowing or estimating the marginal transformations. In addition, the procedures are easy to implement and perform well numerically. The methods are also applied to analyze the Communities and Crime Unnormalized Data from the UCI Machine Learning Repository.

Compared with other methods such as those for the additive regression model and single index model, a significant advantage for our proposed estimation and inference procedures is that they do not require estimation of the marginal transformations. For example, one can select the important variables $x_i$ without any knowledge of the transformations $f_i$. This makes the methods more flexible and adaptive, and achieves the same optimal rate as that for high-dimensional linear regression.

The rest of the paper is organized as follows. After basic notations and definitions are introduced, Section 2 presents the $\ell_1$ penalized minimization procedure for estimating $\boldsymbol{\beta}$ that uses a rank-based correlation matrix estimator. Prediction is also considered. Section 3 constructs a de-biased estimator and establishes an asymptotic normality result. Confidence intervals and hypothesis tests are developed based on the limiting distribution. Numerical performance of the proposed estimator and inference procedures are investigated in Section 4. A brief discussion is given in Section 5 and the main results are proved in Section 6.

# 2 Adaptive Estimation and Prediction

We consider adaptive estimation and prediction in this section. We first introduce the rank-based correlation matrix estimator to extract covariance information on the transformed data that does not require estimation of the marginal transformations, and then present the estimation and prediction procedures and their theoretical properties.

We begin with the basic notations and definitions. Throughout the paper, we use bold-faced letters for vectors. For a vector $\boldsymbol{u} \in \mathbb{R}^p$ and $1 \leq q \leq \infty$, the $\ell_q$ norm is defined as $||\boldsymbol{u}||_q = (\sum_{i=1}^p |u_i|^q)^{1/q}$, with $||\boldsymbol{u}||_\infty = \max_i |u_i|$. In addition, $\boldsymbol{u}[i:j]$ denotes the entries of $\boldsymbol{u}$ from $i$-th to $j$-th coordinates and $\text{supp}(\boldsymbol{u})$ is the support of $\boldsymbol{u}$. For a matrix $A \in \mathbb{R}^{p\times p}$ and $1 \leq q \leq \infty$, the matrix $\ell_q$ operator norm is is defined as $||A||_q = \sup_{||\boldsymbol{u}||_q=1} ||A\boldsymbol{u}||_q$. The spectral norm of $A$ is the $\ell_2$ operator norm and the $\ell_1$ norm is the maximum absolute column sum. For an integer $1 \leq s \leq p$, the $s$-restricted spectral norm of $A$ is defined as $||A||_{2,s} = \sup_{\boldsymbol{u}\in S^{p-1},|\boldsymbol{u}|_0=s} ||A\boldsymbol{u}||_2$, where $S^{p-1}$ is the unit ball in $\mathbb{R}^p$. The vector $\ell_\infty$ norm on matrix $A$ is $|A|_\infty = \max_{i,j} |A_{ij}|$. For a symmetric matrix $A$, we use $\lambda_{\max}(A)$ and $\lambda_{\min}(A)$ to denote respectively the largest and smallest eigenvalue of $A$, and $\kappa(A) = \lambda_{\max}(A)/\lambda_{\min}(A)$ is the condition number. In addition, $\circ$ denotes the matrix element-wise multiplication, and $\otimes$ is the Kronecker product. Moreover, $\text{vec}(\cdot)$ maps an $m \times n$ matrix $A$ to a $\mathbb{R}^{mn}$ vector by laying out the columns of $A$ one by one. For a set of indices $I$,$J$, we let $A_{I,J}$ denote the submatrix formed by the rows in $I$ and columns in $J$. $e_i^{(n)}$ is the $i$-th unit vector in $\mathbb{R}^n$ with entries $e_{ij}^{(n)} = I_{\{j=i\}}$, for $j = 1, ..., n$. $\Phi(\cdot)$ denotes the cumulative distribution function of a standard normal distribution. For two sequences of nonnegative real numbers, $a_n \lesssim b_n$ implies that there exists a constant $C$ not depending on $n$, such that $a_n \le Cb_n$. Finally, we use $[d]$ to denote the set $\{1, 2, ..., d\}$.

### 2.1 Rank-Based Estimator of Correlation Matrix

Recall the model (2), we use $(\boldsymbol{Y}, X)$ to denote the observed data, with $\boldsymbol{Y} \in \mathbb{R}^n$ and $X \in \mathbb{R}^{n\times p}$ the design matrix with rows $\boldsymbol{X}_1^\top, ..., \boldsymbol{X}_n^\top$, and $(\tilde{\boldsymbol{Y}}, \tilde{X})$ to be the original data who possesses the linear relationship. In addition, $\boldsymbol{Z}_i^\top \stackrel{def}{=} (Y_i, \boldsymbol{X}_i^\top)$ and $\tilde{\boldsymbol{Z}}_i^\top \stackrel{def}{=} (\tilde{Y}_i, \tilde{\boldsymbol{X}}_i^\top)$. An essential quantity in estimation of $\boldsymbol{\beta}$ and inference for the Gaussian copula regression model (2) is the covariance matrix (or correlation matrix as the diagonal is 1) $\Sigma$ in (2). Since the marginal transformations $f_i$'s are unknown and thus $(\tilde{Y}, \tilde{\boldsymbol{X}})$ are not directly accessible, the conventional sample covariance matrix is not available as an estimate of $\Sigma$. We thus need an alternative method to estimate the covariance/correlation matrix $\Sigma$.

Our approach is to use the rank-based Kendall's tau, which can be well estimated from the observed data $(Y_1, \boldsymbol{X}_1^\top)$, ..., $(Y_n, \boldsymbol{X}_n^\top)$. This estimator is based on the following fact (see Section 3 of [15]). if $\tilde{\boldsymbol{Z}}_i \stackrel{i.i.d.}{\sim} N_d(0, \Sigma)$ with $\Sigma = (\sigma_{jk})_{1\le j,k\le d}$, then

$$\sigma_{jk} = \sin\left(\frac{\pi}{2}\tau_{jk}\right), \tag{3}$$

where $\tau_{jk}$ is called Kendall's tau and defined as

$$\tau_{jk} = \mathbb{E}[\text{sgn}(\tilde{z}_{1j} - \tilde{z}_{2j})\text{sgn}(\tilde{z}_{1k} - \tilde{z}_{2k})], \tag{4}$$

with $\tilde{\boldsymbol{Z}}_i = (\tilde{z}_{i1}, \tilde{z}_{i2}, ..., \tilde{z}_{id})^\top$, $i = 1, 2$, being two independent copies of $N_d(0, \Sigma)$.

Note that $\tau_{jk}$ given in (4) is invariant under strictly increasing marginal transformations. This leads to an estimate of $\tau_{ij}$ based on the observed data $\boldsymbol{Z}_1$, ..., $\boldsymbol{Z}_n$ under the Gaussian copula regression model

$$\begin{aligned}\hat{\tau}_{jk} &= \frac{2}{n(n-1)} \sum_{1\le i_1<i_2\le n} \text{sgn}(\tilde{Z}_{i_1j} - \tilde{Z}_{i_2j})\text{sgn}(\tilde{Z}_{i_1k} - \tilde{Z}_{i_2k}) \\ &= \frac{2}{n(n-1)} \sum_{1\le i_1<i_2\le n} \text{sgn}(Z_{i_1j} - Z_{i_2j})\text{sgn}(Z_{i_1k} - Z_{i_2k}), \quad 1 \le j,k \le d.\end{aligned} \tag{5}$$

Denote by $\hat{T} = (\hat{\tau}_{jk})_{d\times d}$ the Kendall's tau sample correlation matrix, and its population version $T = (\tau_{jk})_{d\times d}$. Let $\boldsymbol{S}_{i,i'} = (\text{sgn}(Z_{i1} - Z_{i'1}), ..., \text{sgn}(Z_{id} - Z_{i'd}))^\top$, then

$$\hat{T} = (\hat{\tau}_{jk})_{d\times d} = \frac{1}{n(n-1)} \sum_{i\neq i'}^{n} \boldsymbol{S}_{i,i'}\boldsymbol{S}_{i,i'}^\top. \tag{6}$$

Based on the Kendall's tau, (3) immediately leads to the following estimator for the correlation matrix $\Sigma$,

$$\hat{\Sigma} = (\hat{\sigma}_{jk})_{d\times d} \quad \text{with} \quad \widehat{\sigma}_{jk} = \sin\left(\frac{\pi}{2}\widehat{\tau}_{jk}\right). \tag{7}$$

We shall divide $\Sigma$ into four sub-matrices, denoted by $\Sigma_{XX}, \Sigma_{XY}, \Sigma_{YX}, \Sigma_{YY}$, and their corresponding Kendall's tau based estimators are $\hat{\Sigma}_{YY}, \hat{\Sigma}_{YX}, \hat{\Sigma}_{XY}, \hat{\Sigma}_{XX}$, with $\hat{\Sigma}_{YX} = \hat{\Sigma}_{XY}^\top$ and $\Sigma_{YX} = \Sigma_{XY}^\top$.

## 2.2 Estimation of $\boldsymbol{\beta}$

We now introduce the procedure for estimating the sparse coefficient vector $\boldsymbol{\beta}$ in (2). If the marginal transformations $f_i$, $i = 0, 1, ..., p$ were given, then $(\tilde{Y}_i, \tilde{\boldsymbol{X}}_i^\top)$ are available and in this case a natural approach to estimating $\boldsymbol{\beta}$ is to use the Lasso estimator

$$\hat{\boldsymbol{\beta}}_{\text{Lasso}} = \underset{\boldsymbol{\beta}\in\mathbb{R}^p}{\arg\min}\{\frac{1}{2n}||\tilde{\boldsymbol{Y}} - \tilde{X}\boldsymbol{\beta}||_2^2 + \lambda||\boldsymbol{\beta}||_1\}.$$

Rewriting the objective function yields

$$\hat{\boldsymbol{\beta}}_{\text{Lasso}} = \underset{\boldsymbol{\beta}\in\mathbb{R}^p}{\arg\min}\{\frac{1}{2n}(\boldsymbol{\beta}^\top\tilde{X}^\top\tilde{X}\boldsymbol{\beta} - 2\tilde{\boldsymbol{Y}}^\top\tilde{X}) + \lambda||\boldsymbol{\beta}||_1\}. \tag{8}$$

Since $(\tilde{Y}_i, \tilde{\boldsymbol{X}}_i)$ are not directly accessible as the transformations $f_i$'s are unknown, the estimator given in (8) cannot be used. The quantities $\tilde{X}^\top\tilde{X}/n$ and $\tilde{\boldsymbol{Y}}^\top\tilde{X}/n$ in (8) can be viewed as estimators of the covariances $\Sigma_{XX}$ and $\Sigma_{YX}$ respectively. From this perspective, it is natural to replace $\tilde{X}^\top \tilde{X}/n$ and $\tilde{\boldsymbol{Y}}^\top \tilde{X}/n$ in (8) by the alternative covariance estimators $\hat{\Sigma}_{XX}$ and $\hat{\Sigma}_{YX}$ based on Kendall's $\tau$ as discussed in Section 2.1. We thus propose the following $\ell_1$ penalized minimization procedure for estimating $\boldsymbol{\beta}$.

**Algorithm 1:** Adaptive estimator of $\boldsymbol{\beta}$

**Input:** Observed pairs $(Y_1, \boldsymbol{X}_1^\top), ..., (Y_n, \boldsymbol{X}_n^\top)$, parameter $\lambda > 0$.

**Output:** Regularized estimator $\hat{\boldsymbol{\beta}}(\lambda)$.

1: Construct Kendall's tau based covariance estimators $\hat{\Sigma}_{XX}$ and $\hat{\Sigma}_{XY}$.

2: Set

$$\hat{\boldsymbol{\beta}}(\lambda) = \underset{\boldsymbol{\beta} \in \mathbb{R}^p}{\arg\min} \{ \frac{1}{2} (\boldsymbol{\beta}^\top \hat{\Sigma}_{XX} \boldsymbol{\beta} - 2\hat{\Sigma}_{YX} \boldsymbol{\beta}) + \lambda ||\boldsymbol{\beta}||_1 \}. \quad (9)$$

We now consider the properties of the estimator $\hat{\boldsymbol{\beta}}(\lambda)$ given in Algorithm 1. We first define the Restricted Strong Convexity (RSC) condition introduced in [25].

**Definition 1** (RSC). *For a given sparsity level $s \le p$ and constant $\alpha \ge 1$, define the set $C(s, \alpha) := \{\boldsymbol{\theta} \in \mathbb{R}^p : ||\boldsymbol{\theta}_{S^c}||_1 \le \alpha ||\boldsymbol{\theta}_S||_1, S \subset \{1, ..., p\}, |S| \le s\}$. We say a matrix $\Sigma \in \mathbb{R}^{p \times p}$ satisfies the restricted strong convexity (RSC) condition with constants $(\gamma_1, s, \alpha)$, if*

$$\boldsymbol{\theta}^\top \Sigma \boldsymbol{\theta} \ge \gamma_1 ||\boldsymbol{\theta}||_2^2 \quad \text{for all } \boldsymbol{\theta} \in C(s; \alpha).$$

The RSC condition is related to the restricted eigenvalue condition [2] used in the analysis of high-dimensional linear regression. See [25] for more detailed discussion on the RSC.

**Theorem 2.1.** *Assume that $\boldsymbol{\beta}$ is s-sparse. Suppose that $\kappa(\Sigma) \le M$ for some $M > 0$, and $\Sigma_{XX}$ satisfies the RSC with constants $(\gamma_1, s, 3)$. Let $\hat{\boldsymbol{\beta}}(\lambda)$ be defined as (9). If $s = o(\frac{n}{\log p})$, and the tuning parameter $\lambda = C_1 \sqrt{\frac{\log p}{n}}$ is chosen with $C_1 > 2M$, then with probability at least $1 - 2p^{-1}$,*

$$||\hat{\boldsymbol{\beta}}(\lambda) - \boldsymbol{\beta}||_2 \lesssim \sqrt{\frac{s \log p}{n}} \quad \text{and} \quad ||\hat{\boldsymbol{\beta}}(\lambda) - \boldsymbol{\beta}||_1 \lesssim s \sqrt{\frac{\log p}{n}}. \quad (10)$$

*Furthermore, if $|\Sigma_{X_S, X_{S^c}}|_\infty \le 1 - \alpha$ for some constant $\alpha > 0$, where $S = \text{supp}(\boldsymbol{\beta})$ and $X_S$ is its corresponding index set in $\Sigma$, $\min_{i \in S} |\beta_i| \ge \frac{8M}{\gamma_1}(1 + \frac{4(2-\alpha)}{\alpha})\sqrt{\frac{s \log p}{n}}$, then for $\lambda = \frac{8M(2-\alpha)}{\alpha}\sqrt{\frac{s \log p}{n}}$, with probability at least $1 - 2p^{-1}$,*

$$\text{sgn}(\hat{\boldsymbol{\beta}}) = \text{sgn}(\hat{\boldsymbol{\beta}}(\lambda)). \quad (11)$$

The convergence rates of $\hat{\boldsymbol{\beta}}(\lambda)$ under the $\ell_1$ and $\ell_2$ norm losses given in (10) match the minimax lower bounds for high-dimensional linear regression [32]. This implies that $\hat{\boldsymbol{\beta}}(\lambda)$ is minimax rate optimal under the Gaussian copula regression model and achieves the same optimal rate attained by the regular Lasso for linear regression. In other words, the proposed procedure is adaptive to the unknown marginal transformations and gains this added flexibility for free in terms of convergence rate. The result given in (11) shows that, under regularity conditions, $\hat{\boldsymbol{\beta}}(\lambda)$ is sign consistent.

### 2.3 Prediction

In addition to estimation of $\boldsymbol{\beta}$, another problem of signifcant practical interest is predicting the response $Y^*$ for a given value of the covariates $\boldsymbol{x}^* = (x_1^*, ..., x_p^*)$ based on the Gaussian copula regression model (2). In the oracle setting where the transformations $f_0, ..., f_p$ and the coefficient vector $\boldsymbol{\beta}$ are known, then the optimal prediction of the response is

$$\mu^* = f_0^{-1}(\sum_{i=1}^{p} f_i(x_i^*)\beta_i).$$

Our goal is to construct a predictor $\hat{\mu}^*$, based only on the observed data $(Y_1, \boldsymbol{X}_1), ..., (Y_n, \boldsymbol{X}_n)$, that is close to the oracle predictor $\mu^*$.

Let $F_0$ be the cumulative distribution function of $Y$ and let $F_i$ be the cumulative distribution function of $X_i$ for $i = 1, ..., p$. As for the sample version, let $\hat{F}_0$ be the empirical cumulative distribution function of $\{Y_1, ..., Y_n\}$ and let $\hat{F}_i$ be the empirical cumulative distribution function of $\{X_{i1}, ..., X_{in}\}$ for $i = 1, ..., p$. Set

$$\hat{f}_i(t) = \Phi^{-1}(\hat{F}_i(t)). \tag{12}$$

For a given value of the covariates $x^* = (x_1^*, ..., x_p^*)$, we define the predictor

$$\hat{\mu}^* = \hat{f}_0^{-1}(\sum_{i=1}^{p} \hat{f}_i(x_i^*)\hat{\beta}(\lambda)_i), \tag{13}$$

where $\hat{\boldsymbol{\beta}}(\lambda)$ is the estimator given in (9) and $\hat{f}_0^{-1}$ is the generalized inverse of $\hat{f}_0$:

$$\hat{f}_0^{-1}(t) = \inf\{x \in \mathbb{R} : \hat{f}_0(x) \geq t\}.$$

We have the following result for the predictor $\hat{\mu}^*$.

**Theorem 2.2.** *Suppose the conditions in Theorem 2.1 hold. Suppose for some constant* $c > 0$, $|f_0(v_1) - f_0(v_2)| \geq c|v_1 - v_2|$ *for all* $v_1, v_2 \in \mathbb{R}$, *and* $\max_{i=1,\ldots,p} F_i(x_i^*) \in (\delta^*, 1-\delta^*)$ *for some constant* $\delta^* > 0$. *If* $s = o(\sqrt{\frac{n}{\log p}})$, *then the predictor* $\hat{\mu}^*$ *given in* (13) *satisfies, with probability at least* $1 - p^{-1} - n^{-1}$,

$$|\hat{\mu}^* - \mu^*| \lesssim s\sqrt{\frac{\log p}{n}}.$$

This error bound is tight. $f_0(\mu^*) = \sum_{i=1}^p f_i(x_i^*)\beta_i$ can be viewed as a linear functional of $\boldsymbol{\beta}$ with unknown weights $f_i(x_i^*)$ (as the marginal transformations $f_i$ are unknown). For high-dimensional linear regression, inference on the linear functionals of $\boldsymbol{\beta}$ with known weights has been considered in [5], where a lower bound of order $s\sqrt{\frac{\log p}{n}}$ was established for estimation error and for the expected length of confidence intervals for linear functionals with "dense" weight vectors.

# 3 Statistical Inference

We turn in this section to statistical inference for the Gaussian copula regression model. The Lasso estimator is inherently biased as it is essential to trade variance and bias in order to achieve the optimal estimation performance. For statistical inference such as confidence intervals and hypothesis tests, it is desirable to use (nearly) unbiased pivotal estimators. Such an approach has been used in the construction of confidence intervals for high-dimensional linear regression in the recentliterature. See, for example, [5, 13, 37, 43]. We follow the same principle to de-bias the estimator $\hat{\boldsymbol{\beta}}(\lambda)$ given in Algorithm 1.

We begin by noting that $\hat{\boldsymbol{\beta}}(\lambda)$ satisfies the Karush-Kuhn-Tucker (KKT) condition

$$\hat{\Sigma}_{XX}\hat{\boldsymbol{\beta}}(\lambda) - \hat{\Sigma}_{XY} + \lambda\partial||\hat{\boldsymbol{\beta}}(\lambda)||_1 = 0, \tag{14}$$

where $\partial||\hat{\boldsymbol{\beta}}(\lambda)||_1$ is the subgradient of the $\ell_1$ norm $||\cdot||_1$. Equation (14) can be rewritten as

$$\hat{\Sigma}_{XX}(\hat{\boldsymbol{\beta}}(\lambda) - \boldsymbol{\beta}) + \lambda\partial||\hat{\boldsymbol{\beta}}(\lambda)||_1 = \hat{\Sigma}_{XY} - \hat{\Sigma}_{XX}\boldsymbol{\beta}.$$

Suppose one has a good approximation of the "inverse" of $\hat{\Sigma}_{XX}$, say $M$, then

$$M\hat{\Sigma}_{XX}(\hat{\boldsymbol{\beta}}(\lambda) - \boldsymbol{\beta}) + \lambda M\partial||\hat{\boldsymbol{\beta}}(\lambda)||_1 = M(\hat{\Sigma}_{XY} - \hat{\Sigma}_{XX}\boldsymbol{\beta}),$$

and it follows

$$(\hat{\boldsymbol{\beta}}(\lambda) + \lambda M \partial ||\hat{\boldsymbol{\beta}}(\lambda)||_1) - \boldsymbol{\beta} = M(\hat{\Sigma}_{XY} - \hat{\Sigma}_{XX}\boldsymbol{\beta}) + (I - M\hat{\Sigma}_{XX})(\hat{\boldsymbol{\beta}}(\lambda) - \boldsymbol{\beta}), \tag{15}$$

where $(I - M\hat{\Sigma}_{XX})(\hat{\boldsymbol{\beta}}(\lambda) - \boldsymbol{\beta})$ is negligible under mild conditions. This analysis suggests the following de-biasing procedure:

$$\hat{\boldsymbol{\beta}}^u = \hat{\boldsymbol{\beta}}(\lambda) + \lambda M \partial ||\hat{\boldsymbol{\beta}}(\lambda)||_1 = \hat{\boldsymbol{\beta}}(\lambda) + M(\hat{\Sigma}_{XY} - \hat{\Sigma}_{XX}\hat{\boldsymbol{\beta}}(\lambda)),$$

where the second equality is from (14).

We need to construct the matrix $M$ that is a good approximation of the "inverse" of $\hat{\Sigma}_{XX}$. We proceed with two objectives in mind: One is to control $|M\hat{\Sigma}_{XX}|_\infty$ and another is to control the variance of $\hat{\beta}_i^u$. The latter is for the precision of the statistical inference procedures. For example, the length of the confidence intervals for $\beta_i$ is proportional to the standard deviation of $\hat{\beta}_i^u$. Assuming that $(I - M\hat{\Sigma}_{XX})(\hat{\boldsymbol{\beta}}(\lambda) - \boldsymbol{\beta})$ is negligible, the variance of $\hat{\beta}_i^u$ is determined by that of $\boldsymbol{m}_i^\top(\hat{\Sigma}_{XY} - \hat{\Sigma}_{XX}\boldsymbol{\beta})$, where $\boldsymbol{m}_i$ is the $i$-th column of $M$. Let $\boldsymbol{u}_i = (0, \boldsymbol{m}_i^\top)^\top$ and $\boldsymbol{v}_0 = (1, -\boldsymbol{\beta}^\top)^\top \in \mathbb{R}^d$,

$$\boldsymbol{m}_i^\top(\hat{\Sigma}_{XY} - \hat{\Sigma}_{XX}\boldsymbol{\beta}) = \boldsymbol{u}_i\hat{\Sigma}\boldsymbol{v}_0^\top.$$

It will be shown in Lemma 6.3 in Section 6 that the asymptotic variance of $\sqrt{n}\boldsymbol{u}_i\hat{\Sigma}\boldsymbol{v}_0^\top$ is

$$\pi^2\sigma^2_{g1(\boldsymbol{u}_i)} \stackrel{def}{=} \pi^2 \mathrm{Var}(g_1(\boldsymbol{Z}; \boldsymbol{u}_i)), \tag{16}$$

where $g_1(\boldsymbol{Z}; \boldsymbol{u}_i) = \mathbb{E}[g(\boldsymbol{Z}, \boldsymbol{Z}'; \boldsymbol{u}_i)|\boldsymbol{Z}]$, and $g(\boldsymbol{Z}, \boldsymbol{Z}'; \boldsymbol{u}_i)$ is defined as

$$g(\boldsymbol{Z}, \boldsymbol{Z}'; \boldsymbol{u}_i) = \mathrm{sgn}(\boldsymbol{Z} - \boldsymbol{Z}')^\top(\boldsymbol{u}_i\boldsymbol{v}_0^\top \circ \cos(\frac{\pi}{2}T))\mathrm{sgn}(\boldsymbol{Z} - \boldsymbol{Z}')$$

for $\boldsymbol{Z}, \boldsymbol{Z}' \stackrel{i.i.d}{\sim} N_d(0, \Sigma)$ and $\boldsymbol{u}_i \in \mathbb{R}^d$. We need a good estimate of $\sigma^2_{g1(\boldsymbol{u}_i)}$. Note that (16) can be further expressed as

$$\sigma^2_{g1(\boldsymbol{u}_i)} = \mathrm{Var}(g_1(\boldsymbol{Z}; \boldsymbol{u}_i)) = \mathrm{vec}(\boldsymbol{u}_i\boldsymbol{v}_0^\top \circ \cos(\frac{\pi}{2}T))^\top \cdot \Sigma_{h_Z} \cdot \mathrm{vec}(\boldsymbol{u}_i\boldsymbol{v}_0^\top \circ \cos(\frac{\pi}{2}T)). \tag{17}$$

Here $\Sigma_{h_Z} = \mathrm{Var}(h_Z(\boldsymbol{Z})) \in \mathbb{R}^{d^2 \times d^2}$ is the covariance matrix of $h_Z(\boldsymbol{Z}) = \mathbb{E}[\mathrm{sgn}(\boldsymbol{Z} - \boldsymbol{Z}') \otimes \mathrm{sgn}(\boldsymbol{Z} - \boldsymbol{Z}')|\boldsymbol{Z}] \in \mathbb{R}^{d^2}$, which can be estimated by

$$\hat{\Sigma}_{h_Z} = \frac{1}{n}\sum_{i=1}^n(\hat{h}_Z(\boldsymbol{Z}_i) - \frac{1}{n}\sum_{i'=1}^n \hat{h}_Z(\boldsymbol{Z}_{i'}))(\hat{h}_Z(\boldsymbol{Z}_i) - \frac{1}{n}\sum_{i'=1}^n \hat{h}_Z(\boldsymbol{Z}_{i'}))^\top,$$

where $\hat{h}_Z(\boldsymbol{Z}_i) = \frac{1}{n-1}\sum_{i'\neq i}^{n} \mathrm{sgn}(\boldsymbol{Z}_i - \boldsymbol{Z}_{i'}) \otimes \mathrm{sgn}(\boldsymbol{Z}_i - \boldsymbol{Z}_{i'})$.

Then a good estimate of $\sigma^2_{g1(\boldsymbol{u}_i)}$ is given by

$$\hat{\sigma}^2_{g1(\boldsymbol{u}_i)} = \mathrm{vec}(\boldsymbol{u}_i\hat{\boldsymbol{v}}^\top \circ \cos(\frac{\pi}{2}\hat{T}))^\top \hat{\Sigma}_{hZ} \mathrm{vec}(\boldsymbol{u}_i\hat{\boldsymbol{v}}^\top \circ \cos(\frac{\pi}{2}\hat{T})),$$

where $\hat{\boldsymbol{v}} = (1, \hat{\boldsymbol{\beta}}(\lambda)^\top)^\top$.

We are ready to present the de-biasing procedure. To simplify the notation, we define $x(\boldsymbol{u}) : \mathbb{R}^d \to \mathbb{R}^{d^2}$, with $x(\boldsymbol{u}) = \mathrm{vec}(\boldsymbol{u}\boldsymbol{v}_0^\top \circ \cos(\frac{\pi}{2}T))$, and $\hat{x}(\boldsymbol{u}) : \mathbb{R}^d \to \mathbb{R}^{d^2}$, with $\hat{x}(\boldsymbol{u}) = \mathrm{vec}(\boldsymbol{u}\hat{\boldsymbol{v}}^\top \circ \cos(\frac{\pi}{2}\hat{T}))$. Then

$$\sigma^2_{g_1(\boldsymbol{u})} = x(\boldsymbol{u})^\top \Sigma_{hZ} x(\boldsymbol{u}) \quad \text{and} \quad \hat{\sigma}^2_{g_1(\boldsymbol{u})} = \hat{x}(\boldsymbol{u})^\top \hat{\Sigma}_{hZ} \hat{x}(\boldsymbol{u}).$$

---

**Algorithm 2:** De-biased estimator of $\boldsymbol{\beta}$

---

**Input:** Observed pairs $(Y_1, \boldsymbol{X}_1^\top), ..., (Y_n, \boldsymbol{X}_n^\top)$, parameters $a \in (0, \frac{1}{12}), b > 0, \mu > 0, \lambda > 0$.

**Output:** De-biased estimator $\hat{\boldsymbol{\beta}}^u$.

1: Construct Kendall's tau based covariance estimators $\hat{\Sigma}_{XY}$ and $\hat{\Sigma}_{XX}$.

2: Let

$$\hat{\boldsymbol{\beta}}(\lambda) = \min_{\boldsymbol{\beta}\in\mathbb{R}^p} \{\frac{1}{2}(\boldsymbol{\beta}^\top \hat{\Sigma}_{XX}\boldsymbol{\beta} - 2\hat{\Sigma}_{YX}\boldsymbol{\beta}) + \lambda||\boldsymbol{\beta}||_1\}. \tag{18}$$

3: **for** $i = 1, 2, \ldots, p$ **do**

4: Let $\boldsymbol{u}_i$ be a solution of

$$\begin{aligned} \underset{\boldsymbol{u}\in\mathbb{R}^p}{\text{minimize}} \quad & \hat{x}(\boldsymbol{u})^\top \hat{\Sigma}_{hZ} \hat{x}(\boldsymbol{u}) \\ \text{subject to} \quad & ||\hat{\Sigma}_{XX}\boldsymbol{u}[2:d] - e_i^{(p)}||_\infty \leq \mu \\ & \boldsymbol{u}[1] = 0 \\ & b^{-1}n^{-a} \leq ||\boldsymbol{u}||_2 \leq ||\boldsymbol{u}||_1 \leq bn^{a/2} \end{aligned} \tag{19}$$

5: Set $M = (\boldsymbol{u}_1[2:p+1], ..., \boldsymbol{u}_p[2:p+1])$. If any of the above problems is not feasible, then set $M = I_{p\times p}$.

6: Define $\hat{\boldsymbol{\beta}}^u$ as

$$\hat{\boldsymbol{\beta}}^u = \hat{\boldsymbol{\beta}}(\lambda) + M(\hat{\Sigma}_{XY} - \hat{\Sigma}_{XX}\hat{\boldsymbol{\beta}}(\lambda)). \tag{20}$$

---

Note that (19) is a convex program and can be solved efficiently. Let $K = \cos(\frac{\pi}{2}\hat{T}) = (\boldsymbol{K}_1, ..., \boldsymbol{K}_d)$, $\check{\boldsymbol{u}} = (\boldsymbol{u}^\top, ..., \boldsymbol{u}^\top)^\top \in \mathbb{R}^{d^2}$, and $\check{D} = \mathrm{diag}(v_1\,\mathrm{diag}(\boldsymbol{K}_1), ..., v_d\,\mathrm{diag}(\boldsymbol{K}_d))$. Then $\hat{\sigma}^2_{g_1(\boldsymbol{u})}$

can be rewritten as

$$\hat{\sigma}^2_{g_1(\boldsymbol{u})} = \hat{x}(\boldsymbol{u})^\top \hat{\Sigma}_{h_Z} \hat{x}(\boldsymbol{u}) = \check{\boldsymbol{u}}^\top \check{D} \hat{\Sigma}_{h_Z} \check{D} \check{\boldsymbol{u}}. \tag{21}$$

Hence $\hat{\sigma}^2_{g_1(\boldsymbol{u})}$ is convex with respect to $\boldsymbol{u}$. Since the constraints of (19) are a convex set of $\boldsymbol{u}$, these two facts together imply that (19) is a convex program. Note that the first constraint in (19) is to make sure that $M$ is a good approximation of $\hat{\Sigma}^{-1}_{XX}$, and the third constraint is for the convenience of theoretical analysis, in practice $b$ can be chosen sufficiently large so that it does not affect the numerical performance of the algorithm.

The following theorem states the distributional property of $\hat{\boldsymbol{\beta}}^u$ that will serve as the basis for the construction of statistical inference procedures.

**Theorem 3.1.** *Suppose for some constants $M_i > 0$, $i = 1, 2, 3$, that $\frac{1}{M_1} \leq \lambda_{\min}(\Sigma) \leq \lambda_{\max}(\Sigma) \leq M_1$, $||\Sigma^{-1}||_1 < M_2$, and $\lambda_{\min}(\Sigma_{h_Z}) > M_3$. Suppose $s = o(\frac{\sqrt{n}}{\log p})$ and $\mu = a\sqrt{\frac{\log p}{n}}$, and $\lambda = c\sqrt{\frac{\log p}{n}}$ in Algorithm 2 are chosen with $a > 4M_2$ and $c > 2M_1^2$. Then for any fixed $1 \leq i \leq p$ and for all $x \in \mathbb{R}$,*

$$\lim_{n\to\infty} \sup_{\boldsymbol{\beta}\in\mathbb{R}^{p-1}, ||\boldsymbol{\beta}||_0 \leq s} \left| P\left( \frac{\sqrt{n}(\hat{\boldsymbol{\beta}}^u_i - \boldsymbol{\beta}_i)}{\pi\sqrt{\hat{x}(\boldsymbol{u}_i)^\top \hat{\Sigma}_{h_Z} \hat{x}(\boldsymbol{u}_i)}} \leq x \right) - \Phi(x) \right| = 0. \tag{22}$$

Theorem 3.1 shows that the estimator $\hat{\boldsymbol{\beta}}^u$ possesses the similar distributional property as that of the de-biased Lasso estimator in [13], although the observed data here have a linear relationship only after unknown transformations.

The asymptotic normality result given in (22) can be used to construct confidence intervals and hypothesis tests for any given coordinate $\beta_i$. Let $z_{\alpha/2} = \Phi^{-1}(1 - \alpha/2)$.

**Corollary 3.1.** *Suppose the conditions of Theorem 3.1 hold. Then for any given $1 \leq i \leq p$,*

$$CI_i = \left[ \beta^u_i - z_{\alpha/2}\pi\sqrt{\frac{\hat{x}(\boldsymbol{u}_i)^\top \hat{\Sigma}_{h_Z} \hat{x}(\boldsymbol{u}_i)}{n}}, \quad \beta^u_i + z_{\alpha/2}\pi\sqrt{\frac{\hat{x}(\boldsymbol{u}_i)^\top \hat{\Sigma}_{h_Z} \hat{x}(\boldsymbol{u}_i)}{n}} \right] \tag{23}$$

*is an asymptotically $(1 - \alpha)$ level confidence interval for $\beta_i$.*

It is of practical interest to test whether a given covariate $X_i$ is related to the response $Y$. In the context of the Gaussian copula regression model, this can be formulated as testing an individual null hypothesis $H_{0,i} : \beta_i = 0$ versus the alternative $H_{1,i} : \beta_i \neq 0$. To test $H_{0,i}$ against $H_{1,i}$ at the nominal level $\alpha$ for some $0 < \alpha < 1$, based on the asymptotic normality result given in Theorem 3.1, we introduce the test

$$\hat{\Psi}_i = I\left(\frac{\sqrt{n}|\hat{\beta}_i^u|}{\pi\sqrt{\hat{x}(\boldsymbol{u}_i)^\top \hat{\Sigma}_{h_Z} \hat{x}(\boldsymbol{u}_i)}} > z_{\alpha/2}\right). \tag{24}$$

Let $\Psi_i$ be any test for testing $H_{0,i} : \beta_i = 0$ versus $H_{1,i} : \beta_i \neq 0$. Define $\alpha_n(\Psi_i)$ be the size of the test over the collection of $s$-sparse vectors, i.e.,

$$\alpha_n(\Psi_i) = \sup\{P_{\boldsymbol{\beta}}(\Psi_i = 1) : \boldsymbol{\beta} \in \mathbb{R}^p, ||\boldsymbol{\beta}||_0 \leq s, \boldsymbol{\beta}_i = 0\}.$$

For the power of the test, we consider the collection of $s$-sparse vectors with $|\beta_i| \geq \gamma$ for some given $\gamma > 0$ and define the power

$$\zeta_n(\Psi_i; \gamma) = \inf\{P_{\boldsymbol{\beta}}(\Psi_i = 1) : \boldsymbol{\beta} \in \mathbb{R}^p, ||\boldsymbol{\beta}||_0 \leq s, |\beta_i| \geq \gamma\}.$$

**Corollary 3.2.** *Suppose the conditions of Theorem 3.1 hold. The test $\hat{\Psi}_i$ defined in* (24) *satisfies*

$$\lim_{n\to\infty} \alpha_n(\hat{\Psi}_i) \leq \alpha \quad \text{and} \quad \liminf_{n\to\infty} \frac{\zeta_n(\hat{\Psi}_i; \gamma)}{\zeta_n^*(\gamma)} \geq 1, \tag{25}$$

*where $\zeta_n^*(\gamma) \stackrel{def}{=} G(\alpha, \frac{\sqrt{n}\gamma}{\pi\sigma_{g_1(\boldsymbol{u})}})$ with the function $G(\cdot,\cdot)$ defined by*

$$G(\alpha, u) = 2 - \Phi(z_{\alpha/2} + u) - \Phi(z_{\alpha/2} - u).$$

*for $0 < \alpha < 1$ and $u \in \mathbb{R}^+$.*

Consider the problem of testing an individual null hypothesis $H_{0,i} : \beta_i = 0$ versus the alternative $H_{1,i} : \beta_i \neq 0$ under the linear model

$$\tilde{Y}_i = \tilde{\boldsymbol{X}}_i^\top \boldsymbol{\beta} + \epsilon_i, \quad i = 1, 2, ..., n, \tag{26}$$

with $\tilde{\boldsymbol{X}}_i \stackrel{i.i.d}{\sim} N(0, \Sigma_{XX})$ and $\epsilon_i \sim N(0, \sigma^2)$. As shown in [12], for any test $\Psi_i$, if $\alpha_n(\Psi_i) \leq \alpha$, then

$$\limsup_{n\to\infty} \zeta_n(\Psi_i; \gamma) \leq G(\alpha, \frac{\sqrt{n}\gamma}{\sigma_d}),$$

where

$$\sigma_d = \frac{\sigma}{\sqrt{\sigma_{ii} - \Sigma_{i,S}\Sigma_{S,S}^{-1}\Sigma_{S,i}}}.$$

Hence, our test $\hat{\Psi}_i$ has nearly optimal power in the following sense: it has power at least as large as the power of any other test $\Psi_i$ based on a sample of size $\frac{n}{C_d}$, where the factor $C_d = \frac{\pi\sigma_{g_1(\boldsymbol{u}_i)}}{\sigma_d}$.

The results show that the proposed confidence intervals and hypothesis tests share the similar properties as those optimal procedures for the high-dimensional linear regression. They are more flexible in the sense that they are adaptive to unknown monotone marginal transformations.

# 4 Numerical Performance

The proposed estimation and inference procedures are easy to implement. We investigate in this section the numerical performance of the adaptive estimator (9), and we denote it by $\hat{\boldsymbol{\beta}}_{\text{Copula}}(\boldsymbol{Y}, X)$ in this section, as well as the confidence procedure through simulations. The procedures are also applied to the analysis of the Communities and Crime Unnormalized Data from the UCI Machine Learning Repository.

## 4.1 Simulation Results for Estimation Accuracy

We first consider the performance of the the proposed estimator in Gaussian copula regression model defined in (9) by comparing its Root Mean Square error and model selection error with those of the regular Lasso estimator $\hat{\boldsymbol{\beta}}_{\text{Lasso}}(\boldsymbol{Y}, X)$ that is performed on $(\boldsymbol{Y}, X)$ directly, and the Lasso estimator $\hat{\boldsymbol{\beta}}_{\text{Lasso}}(\tilde{\boldsymbol{Y}}, \tilde{X})$ that is performed on $(\tilde{\boldsymbol{Y}}, \tilde{X})$, in which case we assume the marginal transformations $f_i$ are known and $\tilde{\boldsymbol{Y}}$ is linear on $\tilde{X}$.

The simulation setup is as follows. Three cases, $(n, p, s)$=(150, 50, 10), (300, 100, 20), (400, 400, 20) and (400, 200, 20), are analyzed. In each case, we first generate a random Gaussian matrix $A = (a_{i,j})_{1\le i,j\le d}$ where $d = p+1$ and $a_{i,j} \overset{i.i.d.}{\sim} N(0,1)$, then we make the last $p-s$ columns of $A$ orthogonal to the first column of $A$ via the Gram-Schmidt process, and obtain matrix $B$. We then set the covariance matrix $\Sigma = D^{-1/2}(B^\top B + I)^{-1} D^{-1/2}$, where $D = \text{diag}((B^\top B + I)^{-1})$. By this procedure, we zero out the last $p-s$ entries in the first column of $\Sigma^{-1}$, and guarantee the diagonal of $\Sigma$ to be one. Finally, we generate $n$ $i.i.d$ samples $(\tilde{Y}_i, \tilde{\boldsymbol{X}}_i^\top) \sim N_d(0, \Sigma)$. For each choice of $(n, p, s)$, we consider two settings. In the first setting, we set $Y_i = \exp(\tilde{Y}_i)$, $X_{ij} = 2\tilde{X}_{ij}^5 + 1$ for $j = 1, 2, .., 10$, $X_{ij} = -\exp(\tilde{X}_{ij})$ for $j = 11, 12, .., 30$, except for $X_{i,21} = \Phi(\tilde{X}_{i,21})$, bounded by 0 and 1, while in the second setting we constrain $Y_i \in [0,1]$ and set $Y_i = \Phi(\tilde{Y}_i)$ with $X_{ij}$'s transformed the same way as the first setting.

In each setting, the simulation is repeated $N_{\text{sim}} = 500$ times and the tuning parameter $\lambda$ is selected via 5-fold cross validation. The accuracy of the estimators is measured by the average Root Mean Square error

$$e_{\text{est}} = \frac{1}{N_{\text{sim}}} \sum_{i=1}^{N_{\text{sim}}} ||\hat{\boldsymbol{\beta}} - \boldsymbol{\beta}||_2,$$

and the model selection error

$$e_{\text{selection}} = \frac{1}{N_{\text{sim}}} \sum_{i=1}^{N_{\text{sim}}} (\frac{1}{p} \sum_{j=1}^{p} I(\hat{\beta}_j \neq \beta_j)).$$

The simulation results for the three different estimates $\hat{\boldsymbol{\beta}}_{\text{Copula}}(\boldsymbol{Y}, X)$, $\hat{\boldsymbol{\beta}}_{\text{Lasso}}(\tilde{\boldsymbol{Y}}, \tilde{X})$ and $\hat{\boldsymbol{\beta}}_{\text{Lasso}}(\boldsymbol{Y}, X)$ are summarized in Table 1.

| | $\hat{\boldsymbol{\beta}}_{\text{Copula}}(\boldsymbol{Y}, X)$ | | $\hat{\boldsymbol{\beta}}_{\text{Lasso}}(\tilde{\boldsymbol{Y}}, \tilde{X})$ | | $\hat{\boldsymbol{\beta}}_{\text{Lasso}}(\boldsymbol{Y}, X)$ | |
|---|---|---|---|---|---|---|
| $(n, p, s)$ | $e_{\text{selection}}$ | $e_{\text{est}}$ | $e_{\text{selection}}$ | $e_{\text{est}}$ | $e_{\text{selection}}$ | $e_{\text{est}}$ |
| $(150, 50, 10)_1$ | 0.119 | 1.296 | 0.122 | 1.301 | 0.236 | 1.680 |
| $(150, 50, 10)_2$ | 0.119 | 1.296 | 0.122 | 1.301 | 0.324 | 1.721 |
| $(300, 100, 20)_1$ | 0.121 | 1.698 | 0.116 | 1.666 | 0.247 | 2.306 |
| $(300, 100, 20)_2$ | 0.121 | 1.698 | 0.116 | 1.666 | 0.453 | 4.334 |
| $(400, 200, 20)_1$ | 0.068 | 2.202 | 0.082 | 1.554 | 0.143 | 1.799 |
| $(400, 200, 20)_2$ | 0.068 | 2.202 | 0.082 | 1.554 | 0.395 | 4.712 |
| $(400, 400, 40)_1$ | 0.098 | 2.104 | 0.094 | 2.021 | 0.123 | 2.114 |
| $(400, 400, 40)_2$ | 0.098 | 2.104 | 0.094 | 2.021 | 0.325 | 9.854 |

Table 1: Simulation results for the synthetic data described in Section 4. The results corresponds to model selection error $e_{\text{selection}}$ and estimation error $e_{\text{est}}$ for $\hat{\boldsymbol{\beta}}_{\text{Copula}}(\boldsymbol{Y}, X)$, $\hat{\boldsymbol{\beta}}_{\text{Lasso}}(\tilde{\boldsymbol{Y}}, \tilde{X})$ and $\hat{\boldsymbol{\beta}}_{\text{Lasso}}(\boldsymbol{Y}, X)$. The subscript $i$ $(i = 1, 2)$ in $(n, p, s)_i$ denotes the $i$-th setting of transformations

Table 1 shows that the performance of the proposed estimator $\hat{\boldsymbol{\beta}}_{\text{Copula}}(\boldsymbol{Y}, X)$, which does not require the knowledge of the marginal transformations $f_i$, is as good as the oracle estimator $\hat{\boldsymbol{\beta}}_{\text{Lasso}}(\tilde{\boldsymbol{Y}}, \tilde{X})$, which assumes the full knowledge of the transformations $f_i$. As expected, applying the Lasso estimator directly to the observed data leads to severely problematic model selection and parameter estimation.

### 4.2 Simulation Results for Statistical Inference

We now consider the performance of the proposed confidence interval $CI_i$ for the $i$-th coordinate $\beta_i$ given in (23) based on the observed data $(Y_i, \boldsymbol{X}_i^\top)$ in terms of the coverage probability and expected length. In this section we denote the de-biased estimator in (20) as $\hat{\boldsymbol{\beta}}^u_{\text{Copula}}(\boldsymbol{Y}, X)$. The confidence interval is compared with the confidence interval proposed in [13] based on the transformed data $(Y_i, \boldsymbol{X}_i^\top)$ with de-biased estimator $\hat{\boldsymbol{\beta}}^u_{\text{Lasso}}(\boldsymbol{Y}, X)$, and that of $\hat{\boldsymbol{\beta}}^u_{\text{Lasso}}(\tilde{\boldsymbol{Y}}, \tilde{X})$ on the original data $(\tilde{Y}_i, \tilde{\boldsymbol{X}}_i^\top)$ while assuming the marginal transformations $f_i$ are known.In all simulations we set the significance level $\alpha = 0.05$, and consider three cases: $(n, p, s)$=(150, 50, 10), (300, 100, 20) and (400, 200, 20).

In each setting, the simulation is repeated 500 times. The tuning parameter $\lambda$ are selected via 5-fold cross validation, and $\mu, a, b$ in Algorithm 2 are manually set to be $\frac{1}{2}\sqrt{\frac{\log p}{n}}$, $\frac{1}{13}$ and 10 respectively. We discover that the result is robust with respect to the choice of $\mu$, $a$ and $b$. Recall that the $\boldsymbol{\beta}$ is constructed with first $s$ elements nonzero, we construct the 95% confidence intervals for the nonzero (active) coefficient $\beta_1$.

Table 2 summarizes the empirical coverage probability of the nominal 95% confidence intervals and the corresponding average lengths of $\beta_1$. The results show that the empirical coverage probability of $\hat{\boldsymbol{\beta}}^u_{\text{Copula}}(\boldsymbol{Y}, X)$ is very close to the desired confidence level, while it is problematic to construct confidence intervals based on $\hat{\boldsymbol{\beta}}^u_{\text{Lasso}}(\boldsymbol{Y}, X)$. The desired confidence level for the confidence intervals of an active coefficient is always small when we apply the de-biased Lasso estimator directly to the data. The confidence interval constructed by $\hat{\boldsymbol{\beta}}^u_{\text{Copula}}(\boldsymbol{Y}, X)$ performs as good as that constructed by $\hat{\boldsymbol{\beta}}^u_{\text{Lasso}}(\tilde{\boldsymbol{Y}}, \tilde{X})$, which needs additional information of the transformations. In particular, our method tends to have stable confidence interval lengths, while the length of confidence intervals constructed by $\hat{\boldsymbol{\beta}}^u_{\text{Lasso}}(\boldsymbol{Y}, X)$ varies a lot according to the scale of data.

| | $\mathrm{CI}(\hat{\boldsymbol{\beta}}^u_{\text{Copula}}(\boldsymbol{Y}, X))$ | | $\mathrm{CI}(\hat{\boldsymbol{\beta}}^u_{\text{Lasso}}(\tilde{\boldsymbol{Y}}, \tilde{X}))$ | | $\mathrm{CI}(\hat{\boldsymbol{\beta}}^u_{\text{Lasso}}(\boldsymbol{Y}, X))$ | |
|---|---|---|---|---|---|---|
| $(n, p, s)$ | $l(\beta_1)$ | $C(\beta_1)$ | $l(\beta_1)$ | $C(\beta_1)$ | $l(\beta_1)$ | $C(\beta_1)$ |
| $(150, 50, 10)_1$ | 0.252 | 0.950 | 0.333 | 0.946 | 0.025 | 0.150 |
| $(150, 50, 10)_2$ | 0.252 | 0.950 | 0.333 | 0.946 | 0.02 | 0.020 |
| $(300, 100, 20)_1$ | 0.284 | 0.942 | 0.312 | 0.968 | 0.014 | 0.076 |
| $(300, 100, 20)_2$ | 0.284 | 0.942 | 0.312 | 0.968 | 0.001 | 0.014 |
| $(400, 200, 20)_1$ | 0.263 | 0.958 | 0.282 | 0.942 | 0.016 | 0.082 |
| $(400, 200, 20)_2$ | 0.263 | 0.958 | 0.282 | 0.942 | 0.001 | 0.012 |

Table 2: Simulation results for the synthetic data described in Section 4. The results corresponds to 95% confidence intervals. $C(\beta_1)$ and $l(\beta_1)$ respectively stand for coverage probability and average lengths of the confidence interval for $\beta_1$. The subscript $i$ $(i = 1, 2)$ in $(n, p, s)_i$ denotes the $i$-th setting of transformations.

### 4.3 Analysis of Communities and Crime Unnormalized Data

We now apply our estimation and inference procedures on a real data example. The Communities and Crime Unnormalized Data from the UCI Machine Learning Repository combines socio-economic data from the 1990 Census, law enforcement data from the 1990 Law Enforcement Management and Administration Stats survey, and crime data from the 1995 FBI UCR. This dataset has been analyzed in [4, 31]. In this example, we will focus on explaining the response variable, percentage of women who are divorced, using various community characteristics, such as percentage of population that is African American and percent of people in owner occupied households, as well as law enforcement and crime information, such as percent of officers assigned to drug units. In order to further explore the high-dimensional setting, we use the state-level data of Pennsylvania, whose number of predictors is at least as large as the number of observations.

After removing the variables with NA's and two variables directly related to the response (total and male divorce percentages), the data has 101 observations and 114 predictors. To evaluate the performance of the proposed methods, we randomly split the data into a training set with 70 observations, and a test set with 31 observation. We perform such split 100 times, at each time the proposed model is fitted on the training set and the Root Mean Square Error (RMSE) of the prediction (13) is calculated on the test set. Over the 100 random splits of the data, the average RMSE for our method is 1.38. In comparison, performing the regular Lasso on this dataset yields an average RMSE of 3.28. The predicted values by the proposed estimator and those by the Lasso estimator are plotted against the observed values, in one of the testing dataset, as shown in Figure 1.

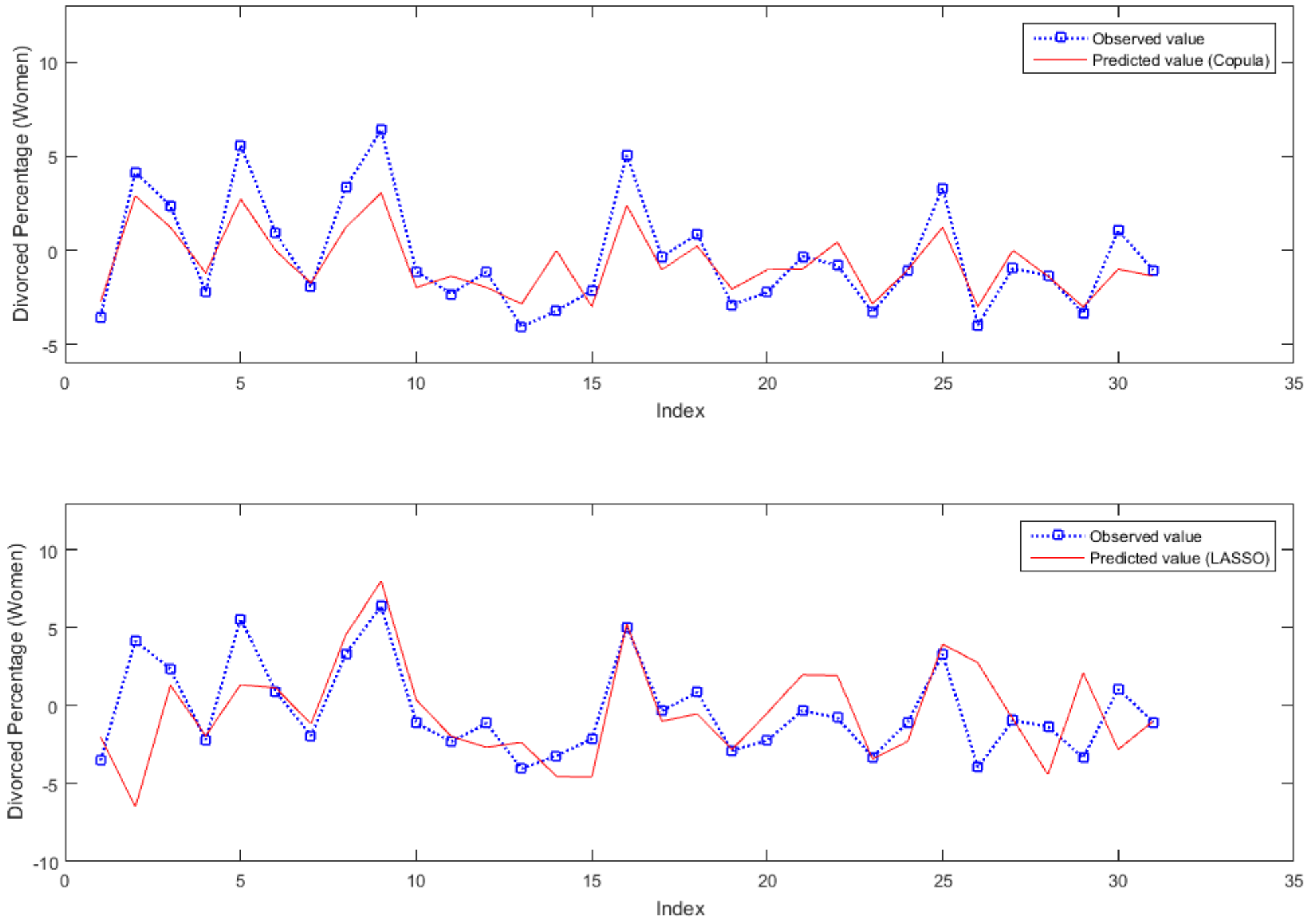

Figure 1: Predicted values by the proposed estimator (top) and Lasso estimator (bottom) are plotted against the observed values in the testing dataset for the divorce percentage of women in the Pennsylvania Communities and Crime Data.

In addition, we use the proposed method for model selection. Applying the procedure to the Communities and Crime Unnormalized Data leads to four selected variables to explain the percentage of women who are divorced: PctFam2Par (percentage of families that are headed by two parents); PctKidsBornNeverMar (percentage of kids born to never married); PctPersOwnOccup (percent of people in owner occupied households) and PctSameHouse85 (percent of people living in the same house as in 1985). This selection procedure correctly exclude all the law enforcement and crime information and irrelevant features in community characteristics, such as the percentage of population that is African American and percentage of people 16 and over who are employed in manufacturing. In addition, the variables selected are all about family/house, which are directly related to divorce percentage.

# 5 Discussion

The Gaussian copula regression model is more flexible than the conventional linear model as it allows for unknown marginal monotonic transformations. The present paper proposes procedures for estimation and statistical inference that are adaptive to the unknown transformations. This is a significant advantage over other methods such as those for the additive regression model and single index model. An important observation is that the objective function for the penalized least squares in classical high-dimensional regression only requires the sample covariances among $X$ and $\boldsymbol{Y}$, which can be replaced by a Kendall's tau based estimator under the Gaussian copula regression model.

This idea can also be generalized to the high-dimensional sparse multivariate regression. For example, under the linear model, the regularized estimator proposed in [34] and the block-structured regularized estimator introduced in [28] only require the knowledge of $X^\top X$ and $X^\top Y$. These can be replaced by the Kendall's tau based estimator $\hat{\Sigma}_{XX}$ and $\hat{\Sigma}_{XY}$ under the Gaussian copula model. Analogous analysis can be carried out to establish estimation consistency and inference results.

Similar ideas can be applied to other related models, such as the additive models in a Reproducing Kernel Hilbert Space (RKHS). In RKHS, the fitting procedure only requires the inner products among data points, and the proposed Algorithm 2 can be modified, via dual representation, for the construction of confidence intervals for additive models in RKHS. In addition, it is also possible to extend the model to discrete data and mixed data, by using the similar idea in [8]. These are interesting topics for future work.

Rank-based correlation matrix estimation has been studied in a number of settings, including the nonparanormal graphical model [1, 17, 39], high dimensional structured covariance/precision matrix estimation [17, 18, 39], and sparse PCA model [10, 24]. In the present paper, we only consider Kendall's tau-based estimator. Alternatively, one may use Spearman's rho. The results are similar and the same technique can be applied.

# 6 Proofs

We prove the main results in this section. We begin by collecting a few technical lemmas that will be used in the proofs of the main results. These lemmas are proved at the end of this section.

## 6.1 Technical Tools

The first lemma shows that the sign vector of a Gaussian random vector is sub-Gaussian.

**Lemma 6.1.** *If $\boldsymbol{Z} \sim N_d(0, \Sigma)$, then $sgn(\boldsymbol{Z}) = (sgn(Z_1), ..., sgn(Z_d))^\top$ is a random vector with subgaussian constant less than $\pi \cdot \kappa(\Sigma)$, that is, for any $\boldsymbol{w} \in S^{d-1}$,*

$$\mathbb{E}[e^{t \cdot \boldsymbol{w}^\top sgn(\boldsymbol{Z})}] \leq e^{t^2 \pi \cdot \kappa(\Sigma)}.$$

The next lemma characterizes the convergence rates of the Kendall's tau based correlation matrix estimator $\hat{\Sigma}$ under different norms.

**Lemma 6.2.** *If $\hat{\Sigma}$ is an estimator of $\Sigma$ based on Kendall's tau, and $\kappa(\Sigma) \leq M$ for some $M > 0$, then*

1. $P(|\hat{\Sigma} - \Sigma|_\infty \lesssim \sqrt{\frac{\log p}{n}}) \geq 1 - 2p^{-2}$;
2. $P(||\hat{\Sigma} - \Sigma||_2 \lesssim \max\{\sqrt{\frac{p+t}{n}}, \frac{p+t}{n}\}) \geq 1 - e^{-t}$;
3. $P(||\hat{\Sigma} - \Sigma||_{2,s} \lesssim \sqrt{\frac{s \log p}{n}}) \geq 1 - p^{-s}$.

Lemma 6.3 below captures the asymptotics of certain $U$-statistics, which will be used to establish the asymptotic results for the proposed estimator.

**Lemma 6.3.** *For $i = 1, 2, ..., p$, let $H_i = \boldsymbol{u}_i[2 : d]^\top (\hat{\Sigma}_{XY} - \hat{\Sigma}_{XX}\boldsymbol{\beta}) = \boldsymbol{u}_i^\top \hat{\Sigma} \boldsymbol{v}_0$, where $\boldsymbol{v}_0 = (1, -\boldsymbol{\beta}^\top)^\top$, then the asymptotic variance of $\sqrt{n} H_i$ is $\pi^2 \sigma^2_{g_1(\boldsymbol{u}_i)}$, and moreover,*

$$\lim_{n \to \infty} \sup_{x \in \mathbb{R}} |P(\frac{\sqrt{n}(H_i - \mathbb{E}[H_i])}{\pi \sigma_{g_1(\boldsymbol{u}_i)}} \leq x) - \Phi(x)| = 0,$$

*where* $\sigma_{g_1(\boldsymbol{u}_i)}$ *is defined in* (17).

Lemmas 6.4, 6.5, 6.6, and 6.7 control the vanishing terms in the construction of confidence intervals for each coordinate $\beta_i$, and all of these four lemmas are stated under the conditions of Theorem 3.1. We use $\boldsymbol{u}$ to denote $\boldsymbol{u}_i$ the solution to (19) for any fixed $i$.

**Lemma 6.4.** *If we take* $\mu = C\sqrt{\frac{\log p}{n}}$ *and* $a, b > 0$ *in Algorithm 2 for large* $C$*, then with probability at least* $1 - 2p^{-2}$*, the optimization problem* (19) *is feasible when* $n$ *is large, that is,*

$$|\Sigma_{XX}^{-1}\hat{\Sigma}_{XX} - I|_\infty \le \mu, \text{ and } b^{-1}n^{-a} \le ||\boldsymbol{u}||_2 \le ||\boldsymbol{u}||_1 \le bn^{a/2}.$$

**Lemma 6.5.** *Let* $\Sigma_{h_Z} = \mathrm{Var}(h_Z(\boldsymbol{Z})) \in \mathbb{R}^{d^2\times d^2}$ *be the covariance matrix of* $h_Z(\boldsymbol{Z}) = \mathbb{E}[sgn(\boldsymbol{Z} - \boldsymbol{Z}') \otimes sgn(\boldsymbol{Z} - \boldsymbol{Z}')|\boldsymbol{Z}]$*, with* $\otimes$ *being the Kronecker product, and its corresponding estimator* $\hat{\Sigma}_{h_Z}$ *is* $\hat{\Sigma}_{h_Z} = \frac{1}{n}\sum_i(\hat{h}_Z(\boldsymbol{Z}_i) - \frac{1}{n}\sum_{i'}\hat{h}_Z(\boldsymbol{Z}_{i'}))(\hat{h}_Z(\boldsymbol{Z}_i) - \frac{1}{n}\sum_{i'}\hat{h}_Z(\boldsymbol{Z}_{i'}))^\top$*, with* $\hat{h}_Z(\boldsymbol{Z}_i) = \frac{1}{n-1}\sum_{i'\neq i} sgn(\boldsymbol{Z}_i - \boldsymbol{Z}_{i'}) \otimes sgn(\boldsymbol{Z}_i - \boldsymbol{Z}_{i'})$*. Then with probability at least* $1 - 5p^{-2}$*,*

$$|x(\boldsymbol{u})^\top(\hat{\Sigma}_{h_Z} - \Sigma_{h_Z})x(\boldsymbol{u})| \lesssim \sqrt{\frac{s\log p}{n^{1-2a}}}.$$

**Lemma 6.6.** *Let* $x(\boldsymbol{u}) = vec(\boldsymbol{u}\boldsymbol{v}_0^\top \circ \cos(\frac{\pi}{2}T))$ *and* $\hat{x}(\boldsymbol{u}) = vec(\boldsymbol{u}\hat{\boldsymbol{v}}^\top \circ \cos(\frac{\pi}{2}\hat{T}))$*, then with probability at least* $1 - p^{-2}$*,*

$$||x(\boldsymbol{u}) - \hat{x}(\boldsymbol{u})||_1 \lesssim n^a\sqrt{\frac{s\log p}{n}}.$$

**Lemma 6.7.** *Let* $\sigma_{g_1(\boldsymbol{u})}$ *be defined as in* (17) *with* $\boldsymbol{u}$ *is the solution to* (19) *with any fixed* $i$*, then*

$$\sigma_{g_1(\boldsymbol{u})}^2 \gtrsim n^{-2a}.$$

The following lemma provides a tight, pointwise deviation inequality of empirical cumulative distribution function, which will be used to establish the consistency of the proposed predictor.

**Lemma 6.8.** *(Adapted from [44]) Let* $\hat{f}_i$ *be defined as* (12) *for* $i \in \{0, 1, , ..., p\}$*, then for any* $\epsilon \in (0, \sqrt{2\pi}]$*, and* $\gamma \in (0, 2)$*, and* $t \in \mathbb{R}$ *such that* $|f_i(t)| \le \sqrt{\gamma \log n}$*, we have*

$$P(|\hat{f}_i(t) - f_i(t)| \ge \epsilon) \le 2\exp(-\frac{n^{1-\gamma/2}}{12\pi\sqrt{2\pi}\sqrt{\gamma\log n}}\epsilon^2) - 3\log(8\pi n^\gamma \log n)\exp(-\frac{1}{64\sqrt{2\pi}}\frac{n^{1-\gamma/2}}{\sqrt{\log n}}),$$

*where* $F_i(t) = \Phi(f_i(t))$.

## 6.2 Proof of Theorem 2.1

This proof relies on the Corollary 1 in [25] and Theorem 3.4 in [16]:

**Lemma 6.9.** *(An adapted version of Corollary 1 in [25] ) If the loss function*

$$L(\boldsymbol{\beta}) = \boldsymbol{\beta}^\top \hat{\Sigma}_{XX}\boldsymbol{\beta} - 2\hat{\Sigma}_{YX}\boldsymbol{\beta} + 1$$

*satisfies restricted strong convexity (RSC), that is*

$$\delta L(\Delta, \boldsymbol{\beta}) \overset{def}{=} L(\boldsymbol{\beta} + \Delta) - L(\boldsymbol{\beta}) - \langle \nabla L(\boldsymbol{\beta}), \Delta \rangle \geq \kappa_L ||\Delta||_2^2, \tag{27}$$

*for some $\kappa_L > 0$ and $\Delta \in C(s) := \{\boldsymbol{\theta} \in \mathbb{R}^p : ||\boldsymbol{\theta}_{S^c}||_1 \leq 3||\boldsymbol{\theta}_S||_1, |S| \leq s\}$.*

*Then for $\lambda \geq 2||\nabla L(\boldsymbol{\beta})||_\infty$, any optimal solution $\hat{\boldsymbol{\beta}}(\lambda)$ to the convex program (9) satisfies the bound*

$$||\hat{\boldsymbol{\beta}}(\lambda) - \boldsymbol{\beta}||_2 \lesssim \sqrt{s}\lambda, \quad ||\hat{\boldsymbol{\beta}}(\lambda) - \boldsymbol{\beta}||_1 \lesssim s\lambda.$$

**Lemma 6.10.** *(An adapted version of Theorem 3.4 in [16]) If we further assume $|\Sigma_{X_S X_{S^c}}|_\infty \leq 1 - \alpha$ for some $\alpha > 0$ and $S = \operatorname{supp}(\boldsymbol{\beta})$ and $\min_{i \in S} |\beta_i| \geq \frac{8}{\gamma_1}(1 + \frac{4(2-\alpha)}{\alpha})M\sqrt{\frac{s \log p}{n}}$, then for $\lambda = \frac{8(2-\alpha)}{\alpha}M\sqrt{\frac{s \log p}{n}}$, with probability at least $1 - 2p^{-1}$,*

$$\operatorname{sgn}(\hat{\boldsymbol{\beta}}) = \operatorname{sgn}(\hat{\boldsymbol{\beta}}(\lambda)).$$

Therefore, to prove Theorem 2.1, it is sufficient to verify (27) and calculate $||\nabla L(\boldsymbol{\beta})||_\infty$. We divide these into two steps.

**Step 1**

By the definition of $\delta L(\Delta, \boldsymbol{\beta})$,

$$\begin{aligned}
\delta L(\Delta, \boldsymbol{\beta}) =& L(\boldsymbol{\beta} + \Delta) - L(\boldsymbol{\beta}) - \langle \nabla L(\boldsymbol{\beta}), \Delta \rangle \\
=& \frac{1}{2}(\boldsymbol{\beta} + \Delta)^\top \hat{\Sigma}_{XX}(\boldsymbol{\beta} + \Delta) - \hat{\Sigma}_{YX}(\boldsymbol{\beta} + \Delta) - \frac{1}{2}\boldsymbol{\beta}^\top \hat{\Sigma}_{XX}\boldsymbol{\beta} \\
& + \hat{\Sigma}_{YX}\boldsymbol{\beta} - \Delta^\top(\hat{\Sigma}_{XX}\boldsymbol{\beta} - \hat{\Sigma}_{XY}) \\
=& \frac{1}{2}\Delta^\top \hat{\Sigma}_{XX}\Delta.
\end{aligned}$$

Before proving (27), we state the adapted version of reduction principle from [35].

**Lemma 6.11.** *(The adapted version of Theorem 10 in [35]) Let $\delta \in (0, \frac{1}{5})$ and $k_0 = 3$. Then there exists a constant $C_0$ that is not dependent with $n, p, s$, such that $\tilde{s} = C_0 s$ and let $E(\tilde{s}) = \{\boldsymbol{w} \in \mathbb{R}^p : ||\boldsymbol{w}||_0 = \tilde{s}\}$ for $\tilde{s} < p$ and $E = \mathbb{R}^p$ otherwise. If $\hat{\Sigma}_{XX}$ satisfies*

$$\forall \boldsymbol{w} \in E(\tilde{s}) \quad (1-\delta)||\boldsymbol{w}||_2^2 \leq \boldsymbol{w}^\top \hat{\Sigma}_{XX} \boldsymbol{w} \leq (1+\delta)||\boldsymbol{w}||_2^2. \tag{28}$$

*Then for any $\boldsymbol{w} \in C(s)$,*

$$(1-5\delta)||\boldsymbol{w}||_2^2 \leq \boldsymbol{w}^\top \hat{\Sigma}_{XX} \boldsymbol{w} \leq (1+3\delta)||\boldsymbol{w}||_2^2 \tag{29}$$

The above claim implies that it is sufficient to show, for $\Delta \in E(\tilde{s}) = \{\boldsymbol{w} \in \mathbb{R}^p : ||\boldsymbol{w}||_0 = \tilde{s}\}$ and some $\delta \in (0, 1/5)$,

$$|\Delta^\top \hat{\Sigma}_{XX} \Delta| \geq (1-\delta)||\Delta||^2.$$

Then Lemma 6.2.2 together with the fact that the spectral norm of a submatrix is bounded by the spectral norm of the whole matrix, for $\Delta \in \{\boldsymbol{w} \in \mathbb{R}^p : ||\boldsymbol{w}||_0 = \tilde{s}\}$, with probability at least $1 - p^{-2}$, we have

$$\begin{aligned}
|\Delta^\top \hat{\Sigma}_{XX} \Delta| =& |\Delta^\top \Sigma_{XX} \Delta + \Delta^\top (\hat{\Sigma}_{XX} - \Sigma_{XX}) \Delta| \\
\geq& |\Delta^\top \Sigma_{XX} \Delta| - |\Delta^\top (\hat{\Sigma}_{XX} - \Sigma_{XX}) \Delta| \\
\geq& |\Delta^\top \Sigma_{XX} \Delta| - ||\hat{\Sigma}_{XX} - \Sigma_{XX}||_{2,\tilde{s}} \cdot ||\Delta||_2^2 \\
\geq& |\Delta^\top \Sigma_{XX} \Delta| - \sqrt{\frac{C_0 s \log p}{n}} ||\Delta||_2^2 \\
\geq& \gamma_1 ||\Delta||_2^2 - \sqrt{\frac{C_0 s \log p}{n}} ||\Delta||_2^2.
\end{aligned}$$

Therefore (27) holds when $s \log p / n \to 0$.

**Step 2**:

$$
\begin{aligned}
&||\nabla L(\boldsymbol{\beta})||_\infty = ||\hat{\Sigma}_{XX}\boldsymbol{\beta} - \hat{\Sigma}_{XY}||_\infty = ||\hat{\Sigma}_{XX}\Sigma_{XX}^{-1}\Sigma_{XY} - \hat{\Sigma}_{XY}||_\infty \\
=&||(\hat{\Sigma}_{XX} - \Sigma_{XX})\Sigma_{XX}^{-1}\Sigma_{XY} + \Sigma_{XY} - \hat{\Sigma}_{XY}||_\infty \\
=&||(\hat{\Sigma}_{XX} - \Sigma_{XX})\boldsymbol{\beta} + \Sigma_{XY} - \hat{\Sigma}_{XY}||_\infty \\
\leq&||(\hat{\Sigma} - \Sigma)(1, -\boldsymbol{\beta}^\top)^\top||_\infty \leq |\hat{\Sigma} - \Sigma|_\infty ||(1, -\boldsymbol{\beta}^\top)^\top||_1 \\
\leq&\sqrt{\frac{\log p}{n}} \cdot (1 + ||\boldsymbol{\beta}||_1) \leq \sqrt{\frac{\log p}{n}} \cdot (1 + \sqrt{s}||\boldsymbol{\beta}||_2) \\
=&\sqrt{\frac{\log p}{n}} \cdot (1 + \sqrt{s}||\Sigma_{XX}^{-1}\Sigma_{XY}||_2) \leq \sqrt{\frac{\log p}{n}} \cdot (1 + \sqrt{s}||\Sigma_{XX}^{-1}||_2||\Sigma_{XY}||_2) \\
\leq&\sqrt{\frac{s\log p}{n}} M.
\end{aligned}
$$

Therefore if we choose $\lambda$ such that $\lambda > 2M\sqrt{\frac{s\log p}{n}}$, then we have $\lambda_n \geq 2||\nabla L(\boldsymbol{\beta})||_\infty$. Then it follows from Theorem 6.9 that, when $s\log p/n \to 0$, with probability at least $1 - 2p^{-2}$,

$$
\begin{aligned}
&||\hat{\boldsymbol{\beta}}(\lambda) - \boldsymbol{\beta}||_2 \lesssim \sqrt{s}\lambda \lesssim \sqrt{\frac{s\log p}{n}} \\
&||\hat{\boldsymbol{\beta}}(\lambda) - \boldsymbol{\beta}||_1 \lesssim s\lambda \lesssim s\sqrt{\frac{\log p}{n}} \\
&\operatorname{sgn}(\hat{\boldsymbol{\beta}}_0) = \operatorname{sgn}(\hat{\boldsymbol{\beta}}(\lambda)).
\end{aligned}
$$

□

## 6.3 Proof of Theorem 2.2

According to Lemma 6.8 and by the union bound

$$
\begin{aligned}
P(\max_{i\in[0,1,2,\ldots,p]} |\hat{f}_i(t) - f_i(t)| \geq \epsilon) \leq &2\exp(\log d - \frac{n^{1-\gamma/2}}{12\pi\sqrt{2\pi}\sqrt{\gamma\log n}}\epsilon^2) \\
&- 3\log(8\pi n^\gamma \log n)\exp(\log d - \frac{1}{64\sqrt{2\pi}}\frac{n}{\sqrt{n^\gamma \log n}}).
\end{aligned}
$$

Therefore by taking $\epsilon = \sqrt{\frac{24\pi\sqrt{2\pi}\sqrt{\gamma\log n}\log d}{n^{1-\gamma/2}}}$, then for $t \in \mathbb{R}$ such that $|f_i(t)| \leq \sqrt{\gamma \log n}$, with probability at least $1 - d^{-1} - n^{-1}$,

$$
\max_{i\in[0,1,2,\ldots,p]} |\hat{f}_i(t) - f_i(t)| \lesssim \frac{(\gamma\log n)^{1/4}\sqrt{\log d}}{n^{1/2-\gamma/4}}. \tag{30}
$$

Since $\max_{i=1,\ldots,p} F_i(x_i^*) \in (\delta^*, 1-\delta^*)$, there exists some constant $M_* > 0$, such that

$$\max_{i=1,\ldots,p} f_i(x_i^*) = \max_{i=1,\ldots,p} \Phi^{-1}(F_i(x_i^*)) < M_*.$$

Therefore, if we let $\gamma = \frac{M_*^2}{\log n}$, we have $\max_{i=1,\ldots,p} f_i(x_i^*) \leq \sqrt{\gamma \log n}$. Then by (30), with probability at least $1 - d^{-1} - n^{-1}$,

$$\max_{i\in[0,1,2,\ldots,p]} |\hat{f}_i(x_i^*) - f_i(x_i^*)| \lesssim \sqrt{\frac{\log d}{n}}.$$

Combining the result in Theorem 2.1, with probability at least $1 - 2d^{-1} - n^{-1}$,

$$\begin{aligned}
|y^* - \hat{y}^*| =& |\hat{f}_0^{-1}(\sum_{i=1}^{p} \hat{f}_i(x_i^*)\hat{\beta}(\lambda)_i) - f_0^{-1}(\sum_{i=1}^{p} f_i(x_i^*)\beta(\lambda)_i)| \\
\lesssim& |\sum_{i=1}^{p} \hat{f}_i(x_i^*)\hat{\beta}(\lambda)_i - \sum_{i=1}^{p} f_i(x_i^*)\beta(\lambda)_i| \\
\leq& |\sum_{i=1}^{p} \hat{f}_i(x_i^*)\hat{\beta}(\lambda)_i - \sum_{i=1}^{p} f_i(x_i^*)\hat{\beta}(\lambda)_i| + |\sum_{i=1}^{p} f_i(x_i^*)\hat{\beta}(\lambda)_i - \sum_{i=1}^{p} f_i(x_i^*)\beta(\lambda)_i| \\
\lesssim& (||\boldsymbol{\beta}||_1 + s\sqrt{\frac{\log p}{n}}) \cdot \max_{i\in[0,1,2,\ldots,p]} |\hat{f}_i(t) - f_i(t)| + ||\hat{\boldsymbol{\beta}}(\lambda) - \boldsymbol{\beta}||_1 \\
\leq& ||\hat{\boldsymbol{\beta}}(\lambda) - \boldsymbol{\beta}||_1 + (s||\boldsymbol{\beta}||_2 + s\sqrt{\frac{\log p}{n}}) \cdot \max_{i\in[0,1,2,\ldots,p]} |\hat{f}_i(t) - f_i(t)| \\
\lesssim& s\sqrt{\frac{\log d}{n}},
\end{aligned}$$

where the last inequality results from the fact $\boldsymbol{\beta} = \Sigma_{XX}^{-1}\Sigma_{XY}$, and then

$$||\boldsymbol{\beta}||_2 = ||\Sigma_{XX}^{-1}\Sigma_{XY}||_2 \leq \frac{\lambda_{\max}(\Sigma)}{\lambda_{\min}(\Sigma)} \leq M.$$

What's more, the first inequality is due to the following claim.

**Claim**: For two increasing functions $f_1, f_2$, if $|f_1(f_1^{-1}(t)) - f_2(f_1^{-1}(t))| < c_1$ for some $t \in \mathbb{R}$ and $c_1 > 0$, and $|f_2(v_1) - f_2(v_2)| \geq c_2|v_1 - v_2|$ for some $c_2 > 0$, then

$$|f_1^{-1}(t) - f_2^{-1}(t)| \leq \frac{c_1}{c_2}.$$

In effect, if $|f_1^{-1}(t) - f_2^{-1}(t)| > \frac{c_1}{c_2}$, then

$$\begin{aligned}
|f_1(f_1^{-1}(t)) - f_2(f_1^{-1}(t))| =& |f_1(f_1^{-1}(t)) - f_2(f_2^{-1}(t)) + f_2(f_2^{-1}(t)) - f_2(f_1^{-1}(t))| \\
\geq& |f_2(f_2^{-1}(t)) - f_2(f_1^{-1}(t))| - |f_1(f_1^{-1}(t)) - f_2(f_2^{-1}(t))| \\
>& c_2 \cdot \frac{c_1}{c_2} - 0 = c_1.
\end{aligned}$$

This leads to a contradiction.

$\square$

## 6.4 Proof of Theorem 3.1

Before we proceed, we should determine $\mu$ to make the optimization problem (19) feasible. By Lemma 6.4, it is sufficient to set $\mu = C\sqrt{\frac{\log p}{n}}$ for some sufficient large constant $C$.

According to (20) in Algorithm 2,

$$\begin{aligned}\hat{\boldsymbol{\beta}}^u &= \hat{\boldsymbol{\beta}}(\lambda) + M(\hat{\Sigma}_{XY} - \hat{\Sigma}_{XX}\hat{\boldsymbol{\beta}}(\lambda)) \\ &= \boldsymbol{\beta} - \boldsymbol{\beta} + \hat{\boldsymbol{\beta}}(\lambda) + M\hat{\Sigma}_{XY} - M\hat{\Sigma}_{XX}\hat{\boldsymbol{\beta}}(\lambda) \\ &= \boldsymbol{\beta} + (M\hat{\Sigma}_{XY} - M\hat{\Sigma}_{XX}\boldsymbol{\beta}) + (M\hat{\Sigma}_{XX} - I)(\boldsymbol{\beta} - \hat{\boldsymbol{\beta}}(\lambda)).\end{aligned}$$

This implies

$$\sqrt{n}(\hat{\boldsymbol{\beta}}^u - \boldsymbol{\beta}(\lambda)) = \sqrt{n}(M\hat{\Sigma}_{XY} - M\hat{\Sigma}_{XX}\boldsymbol{\beta}) + \sqrt{n}(I - M\hat{\Sigma}_{XX})(\boldsymbol{\beta} - \hat{\boldsymbol{\beta}}(\lambda)). \tag{31}$$

We control the two terms on the right hand side separately.

**Step 1**: $||\sqrt{n}(I - M\hat{\Sigma}_{XX})(\boldsymbol{\beta} - \hat{\boldsymbol{\beta}}(\lambda))||_\infty \to 0$ with high probability.

By Theorem 2.1 and Lemma 6.4, with probability at least $1 - 3p^{-2}$,

$$\begin{aligned}||\sqrt{n}(I - M\hat{\Sigma}_{XX})(\boldsymbol{\beta} - \hat{\boldsymbol{\beta}}(\lambda))||_\infty \leq& \sqrt{n}||I - M\hat{\Sigma}_{XX}||_\infty ||\boldsymbol{\beta} - \hat{\boldsymbol{\beta}}(\lambda)||_1 \\ \leq& \sqrt{n}\mu \cdot s\sqrt{\frac{\log p}{n}} \lesssim \sqrt{n}\sqrt{\frac{\log p}{n}} \cdot s\sqrt{\frac{\log p}{n}}.\end{aligned}$$

Therefore, when $\frac{s\log p}{\sqrt{n}} \to 0$, with probability at least $1 - 3p^{-2}$,

$$||\sqrt{n}(I - M\hat{\Sigma}_{XX})(\boldsymbol{\beta} - \hat{\boldsymbol{\beta}}(\lambda))||_\infty \to 0.$$

**Step 2**: Asymptotics of $\sqrt{n}(\boldsymbol{u}_i'\hat{\Sigma}_{XY} - \boldsymbol{u}_i'\hat{\Sigma}_{XX}\boldsymbol{\beta})$.

With Lemma 6.5, Lemma 6.6, and by $|\Sigma_{h_Z}|_\infty \leq 1$, when $\frac{s\log p}{\sqrt{n}} \to 0$, we have with probability at least $1 - p^{-2}$,

$$
\begin{aligned}
|\sigma^2_{g_1(\boldsymbol{u}_i)} - \hat{\sigma}^2_{g_1(\boldsymbol{u}_i)}| &= |x(\boldsymbol{u}_i)^\top \Sigma_{h_Z} x(\boldsymbol{u}_i) - \hat{x}(\boldsymbol{u}_i)^\top \hat{\Sigma}_{h_Z} \hat{x}(\boldsymbol{u}_i)| \\
&\leq |(x(\boldsymbol{u}_i) - \hat{x}(\boldsymbol{u}_i))^\top \Sigma_{h_Z}(x(\boldsymbol{u}_i) - \hat{x}(\boldsymbol{u}_i))| + |x(\boldsymbol{u}_i)^\top (\hat{\Sigma}_{h_Z} - \Sigma_{h_Z}) x(\boldsymbol{u}_i)| \\
&\leq ||x(\boldsymbol{u}_i) - \hat{x}(\boldsymbol{u}_i)||_1^2 + |x(\boldsymbol{u}_i)^\top (\hat{\Sigma}_{h_Z} - \Sigma_{h_Z}) x(\boldsymbol{u}_i)| \\
&\lesssim n^{2a} \frac{s \log p}{n} + \sqrt{\frac{s \log p}{n^{1-2a}}} \lesssim \sqrt{\frac{s \log p}{n^{1-2a}}}
\end{aligned}
$$

Lemma 6.7 shows $\sigma^2_{g_1(\boldsymbol{u}_i)} \gtrsim n^{-2a}$. It follows $|\frac{\hat{\sigma}^2_{g_1(\boldsymbol{u}_i)}}{\sigma^2_{g_1(\boldsymbol{u}_i)}} - 1| \lesssim \sqrt{\frac{s \log p}{n^{1-6a}}}$. In addition, due to the positiveness of $\sigma_{g1}$ and $\hat{\sigma}_{g_1}$, when $\frac{s \log p}{\sqrt{n}} \to 0$ and $a < \frac{1}{12}$, $\hat{\sigma}_{g_1(\boldsymbol{u}_i)}/\sigma_{g_1(\boldsymbol{u}_i)} \to 1$ in probability. Then according to Lemma 6.3, for any $\epsilon > 0$,

$$
\begin{aligned}
P(\frac{\sqrt{n}(H_i - \mathbb{E}[H_i])}{\pi \hat{\sigma}_{g_1(\boldsymbol{u}_i)}} \leq x) =& P(\frac{\sigma_{g_1(\boldsymbol{u}_i)}}{\hat{\sigma}_{g_1(\boldsymbol{u}_i)}} \frac{\sqrt{n}(H_i - \mathbb{E}[H_i])}{\pi \sigma_{g_1(\boldsymbol{u}_i)}} \leq x) \\
\leq& P(\frac{\sqrt{n}(H_i - \mathbb{E}[H_i])}{\pi \sigma_{g_1(\boldsymbol{u}_i)}} \leq \frac{x}{1-\epsilon}) + P(\frac{\hat{\sigma}_{g_1(\boldsymbol{u}_i)}}{\sigma_{g_1(\boldsymbol{u}_i)}} \geq \frac{1}{1-\epsilon}) \\
\to& \Phi(\frac{x}{1-\epsilon}) \qquad \text{as } n \to \infty,
\end{aligned}
$$

where the last limit results from Lemma 6.3.

Let $\epsilon \to 0$, we have

$$
\limsup_{n \to \infty} P(\frac{\sqrt{n}(H_i - \mathbb{E}[H_i])}{\pi \hat{\sigma}_{g_1(\boldsymbol{u}_i)}} \leq x) \leq \Phi(x).
$$

Similarly, we have

$$
P(\frac{\sqrt{n}(H_i - \mathbb{E}[H_i])}{\pi \hat{\sigma}_{g_1(\boldsymbol{u}_i)}} \leq x) \geq P(\frac{\sqrt{n}(H_i - \mathbb{E}[H_i])}{\pi \sigma_{g_1(\boldsymbol{u}_i)}} \leq x(1-\epsilon)) - P(\frac{\hat{\sigma}_{g_1(\boldsymbol{u}_i)}}{\sigma_{g_1(\boldsymbol{u}_i)}} \leq 1 - \epsilon)
$$

This leads to

$$
\liminf_{n \to \infty} P(\frac{\sqrt{n}(H_i - \mathbb{E}[H_i])}{\pi \hat{\sigma}_{g_1(\boldsymbol{u}_i)}} \leq x) \geq \Phi(x).
$$

In conclusion, when $\frac{s \log p}{\sqrt{n}} \to 0$, we have

$$
\lim_{n \to \infty} \sup_{x \in \mathbb{R}} |P(\frac{\sqrt{n}(H_i - \mathbb{E}[H_i])}{\pi \hat{\sigma}_{g_1(\boldsymbol{u}_i)}} \leq x) - \Phi(x)| = 0.
$$

□

# 7 Proof of Auxiliary Lemmas

### Proof of Lemma 6.1

Let $A_1, A_2 \in \mathbb{R}^{d\times 2d}$ with each row of $A_i$ has unit norm, and for some diagonal matrix $D = \text{diag}(m_1, m_2, ..., m_d)$, satisfy

$$\begin{cases} A_1 P_{A_2}^{\perp} A_1^{\top} = D \\ (A_1 + A_2)(A_1 + A_2)^{\top} = \Sigma \\ \text{rank}(A_1) = \text{rank}(A_2) = d, \end{cases} \tag{32}$$

where $P_{A_2}^{\perp} = I_{2d} - A_2^{\top}(A_2 A_2^{\top})^{-1} A_2$. Therefore, if $\begin{pmatrix} \boldsymbol{X} \\ \boldsymbol{Y} \end{pmatrix} \sim N(0, \begin{pmatrix} \Sigma_{11} & \Sigma_{12} \\ \Sigma_{21} & \Sigma_{22} \end{pmatrix})$ with $\Sigma_{11} = A_1 A_1^{\top}$, $\Sigma_{22} = A_2 A_2^{\top}$, $\Sigma_{12} = A_1 A_2^{\top}$, then

$$\Sigma_{11\cdot 2} \stackrel{def}{=} \Sigma_{11} - \Sigma_{12}\Sigma_{22}^{-1}\Sigma_{21} = D,$$
$$\Sigma_{11} + \Sigma_{12} + \Sigma_{21} + \Sigma_{22} = \Sigma.$$

This implies $\boldsymbol{X}|\boldsymbol{Y} \sim N(\Sigma_{12}\Sigma_{22}^{-1}\boldsymbol{Y}, D)$, $\boldsymbol{X}+\boldsymbol{Y} \sim N(0, \Sigma)$. For $\boldsymbol{v} \in \mathbb{R}^d$ with $||\boldsymbol{v}||_2 = 1$,

$$\begin{aligned} \mathbb{E}e^{v^{\top}\text{sgn}(\boldsymbol{Z})} &= \mathbb{E}e^{v^{\top}\text{sgn}(\boldsymbol{X}+\boldsymbol{Y})} = \mathbb{E}[\mathbb{E}[e^{v^{\top}\text{sgn}(\boldsymbol{X}+\boldsymbol{Y})}|\boldsymbol{Y}] \\ &= \mathbb{E}[\mathbb{E}[e^{\sum_{i=1}^d v_i \text{sgn}(X_i+Y_i)}|Y] = \mathbb{E}[\prod_{i=1}^d \mathbb{E}[e^{v_i \text{sgn}(X_i+Y_i)}|\boldsymbol{Y}]]. \end{aligned}$$

We have

$$\text{sgn}(X_i + Y_i)|\boldsymbol{Y} \sim \begin{cases} 1, & \text{with probability } \Phi(\frac{Y_i + e_i^{\top}\Sigma_{12}\Sigma_{22}^{-1}\boldsymbol{Y}}{\sqrt{m_i}}), \\ -1, & \text{with probability } 1 - \Phi(\frac{Y_i + e_i^{\top}\Sigma_{12}\Sigma_{22}^{-1}\boldsymbol{Y}}{\sqrt{m_i}}). \end{cases} \tag{33}$$

Let $h_i(\boldsymbol{Y}) \stackrel{def}{=} \mathbb{E}[\text{sgn}(X_i+Y_i)|\boldsymbol{Y}] = 2\Phi(\frac{Y_i + e_i^{\top}\Sigma_{12}\Sigma_{22}^{-1}\boldsymbol{Y}}{\sqrt{m_i}}) - 1 = 2\Phi(\frac{e_i^{\top}(I + \Sigma_{12}\Sigma_{22}^{-1})\boldsymbol{Y}}{\sqrt{m_i}}) - 1$.

Therefore

$$\mathbb{E}e^{v^{\top}\text{sgn}(\boldsymbol{Z})} = \mathbb{E}[\prod_{i=1}^d \mathbb{E}[e^{v_i(\text{sgn}(X_i+Y_i) - h_i(\boldsymbol{Y}))}|\boldsymbol{Y}]e^{v_i h_i(\boldsymbol{Y})}] = e^{1/2}\mathbb{E}[\prod_{i=1}^d e^{v_i h_i(\boldsymbol{Y})}] = e^{1/2}\mathbb{E}[e^{\sum_{i=1}^d v_i h_i(\boldsymbol{Y})}].$$

Let $g(\tilde{\boldsymbol{Y}}) = \sum_{i=1}^d v_i h_i(\Sigma_{22}^{1/2}\tilde{\boldsymbol{Y}})$ where $\tilde{\boldsymbol{Y}} = \Sigma_{22}^{-1/2}\boldsymbol{Y} \sim N_d(0, I)$, then we have

$$\begin{aligned}
|g(\tilde{\boldsymbol{Y}}_1) - g(\tilde{\boldsymbol{Y}}_2)| = |\sum_{i=1}^d v_i(h_i(\Sigma_{22}^{1/2}\tilde{\boldsymbol{Y}}_1) - h_i(\Sigma_{22}^{1/2}\tilde{\boldsymbol{Y}}_2))| &\leq \sqrt{\sum_{i=1}^d (h_i(\Sigma_{22}^{1/2}\tilde{\boldsymbol{Y}}_1) - h_i(\Sigma_{22}^{1/2}\tilde{\boldsymbol{Y}}_2))^2} \\
&\leq \pi||D^{-1/2}(I + \Sigma_{12}\Sigma_{22}^{-1})\Sigma_{22}^{1/2}|| \cdot ||\tilde{\boldsymbol{Y}}_1 - \tilde{\boldsymbol{Y}}_2|| \\
&\leq \pi\sqrt{||D^{-1/2}(I + \Sigma_{12}\Sigma_{22}^{-1})\Sigma_{22}(I + \Sigma_{12}\Sigma_{22}^{-1})^\top D^{-1/2}||} \cdot ||\tilde{\boldsymbol{Y}}_1 - \tilde{\boldsymbol{Y}}_2|| \\
&= \pi\sqrt{||D^{-1/2}\Sigma - 1||} \cdot ||\tilde{\boldsymbol{Y}}_1 - \tilde{\boldsymbol{Y}}_2||.
\end{aligned}$$

From (1) we know that $0 \leq (A_1 + A_2)P_{A_2}^{\perp}(A_1 + A_2)^\top = \Sigma - D$. Then we have

$$|g(\tilde{\boldsymbol{Y}}_1) - g(\tilde{\boldsymbol{Y}}_2)| \leq \frac{\lambda_{\max}(\Sigma) - \lambda_{\min}(\Sigma)}{\lambda_{\min}(\Sigma)}\pi||\tilde{\boldsymbol{Y}}_1 - \tilde{\boldsymbol{Y}}_2||.$$

Thus, the Lipschitz norm of $g(\cdot)$ is bounded by $\frac{\lambda_{\max}(\Sigma) - \lambda_{\min}(\Sigma)}{\lambda_{\min}(\Sigma)}\pi$. By the Gaussian concentration inequality [3], $\mathbb{E}[e^{\sum_{i=1}^d v_i h_i(\boldsymbol{Y})}] = e^{\pi}\mathbb{E}[e^{g(\Sigma_{22}^{-1/2}\boldsymbol{Y})}] \leq e^{\pi M/2}$ with $M = ||D^{-1}\Sigma||_2$. If we let $D = \lambda_{\min}(\Sigma)I$, then $M = \kappa(\Sigma)$. □

## Proof of Lemma 6.2

1. According to Taylor's expansion, $|\hat{\Sigma} - \Sigma|_\infty$ can be bounded by $|\hat{T} - T|_\infty$:

$$\begin{aligned}
\hat{\Sigma} - \Sigma &= \sin(\frac{\pi}{2}\hat{T}) - \sin(\frac{\pi}{2}T) \\
&= \cos(\frac{\pi}{2}T) \circ (\hat{T} - T) \cdot \frac{\pi}{2} - \frac{1}{2}\sin(\frac{\pi}{2}) \circ \frac{\pi}{2}(\hat{T} - T) \circ \frac{\pi}{2}(\hat{T} - T).
\end{aligned}$$

This implies $|\hat{\Sigma} - \Sigma|_\infty \lesssim |\hat{T} - T|_\infty + |\hat{T} - T|_\infty^2$. By Hoeffding inequality, $P(|T_{jk} - T_{jk}| > t) \leq 2\exp(-nt^2/4)$. Therefore,

$$\begin{aligned}
P(|\hat{T} - T|_\infty > t) &\leq \sum_{j,k=1}^p P(|T_{jk} - T_{jk}| > t) \\
&\leq 2p^2\exp(-nt^2/4) = 2\exp(2\log p - nt^2/4).
\end{aligned}$$

Let $t = 4\sqrt{\frac{\log p}{n}}$, the above inequality implies that with probability $1 - 2p^{-2}$,

$$|\hat{T} - T|_\infty \lesssim \sqrt{\frac{\log p}{n}}.$$

This shows that with probability $1 - 2p^{-2}$,

$$|\hat{\Sigma} - \Sigma|_\infty \leq |\hat{T} - T|_\infty + |\hat{T} - T|_\infty^2 \lesssim \sqrt{\frac{\log p}{n}}.$$

2. Let $d = p+1$, and without loss of generality we assume $n$ is even. For $i, i' \in \{1, 2, ..., n\}$, define $\boldsymbol{S}_{i,i'} = \mathrm{sgn}(\boldsymbol{Z}_i - \boldsymbol{Z}_{i'}) = (\mathrm{sgn}(Z_{i1} - Z_{i'1}), ..., \mathrm{sgn}(Z_{id} - Z_{i'd}))^\top$, and $\hat{\Delta}_{i,i'} = \frac{1}{n(n-1)}(\boldsymbol{S}_{i,i'}\boldsymbol{S}_{i,i'}^\top - T)$. Moreover, for any permutation $\sigma \in S_n$, where $S_n$ is the permutation group of $\{1, ..., n\}$, let $(i_1, ..., i_n) = \sigma(1, ..., n)$. For $r = 1, ..., n/2$ (without loss of generality, we assume $n$ is even), we define $\boldsymbol{S}_r^\sigma$ and $\hat{\Delta}_r^\sigma$ to be $\boldsymbol{S}_r^\sigma = \boldsymbol{S}_{2i_r, 2i_r - 1}$, $\hat{\Delta}_r^\sigma = \frac{1}{n/2}(\boldsymbol{S}_r^\sigma \boldsymbol{S}_r^{\sigma T} - T)$. Then

$$\Delta = \hat{T} - T = \sum_{i,i'} \Delta_{i,i'} = \frac{1}{|S_n|} \sum_{\sigma \in S_n} \sum_{r=1}^{n/2} \hat{\Delta}_r^\sigma.$$

and consequently,

$$||\hat{T} - T|| \leq \frac{1}{|S_n|} \sum_{\sigma \in S_n} \sum_{r=1}^{n/2} \hat{\Delta}_r^\sigma.$$

Let $N_\epsilon$ be the largest number of $\epsilon$-balls one can pack in the $(1 + \epsilon)$-ball centered at the origin and $\{\boldsymbol{w}_{(j)}, j \leq N_\epsilon\}$ be the centers of such $\epsilon$-balls in one of such configurations. From straight forward volume comparison we have

$$N_\epsilon \leq (1/\epsilon + 1)^d.$$

For each $\boldsymbol{w} \in S^{d-1}$, $||\boldsymbol{w} - \boldsymbol{w}_{(j)}||_2 \leq 2\epsilon$ for some $j \leq N_\epsilon$, so that

$$\begin{aligned} |\boldsymbol{w}^\top \Delta \boldsymbol{w}| \leq & |\boldsymbol{w}_{(j)}^\top \Delta \boldsymbol{w}_{(j)}| + |(\boldsymbol{w} - \boldsymbol{w}_{(j)})^\top \Delta (\boldsymbol{w} - \boldsymbol{w}_{(j)})| \\ \leq & |\boldsymbol{w}_{(j)}^\top \Delta \boldsymbol{w}_{(j)}| + 4\epsilon^2 ||\Delta||_2. \end{aligned}$$

This implies

$$||\Delta||_2 \leq \sup_{j \leq N_\epsilon} \frac{|\boldsymbol{w}_{(j)}^\top \Delta \boldsymbol{w}_{(j)}|}{1 - \epsilon^2}, \tag{34}$$

with $N_\epsilon \leq (1 + 1/\epsilon)^d$.

In addition, for any $\boldsymbol{w} \in S^{d-1}$, according to Lemma 6.1, we have

$$\mathbb{E}[e^{t\boldsymbol{w}^\top (\sum_{r=1}^{n/2} \hat{\Delta}_r^\sigma)\boldsymbol{w}}] = \prod_{r=1}^{n/2} \mathbb{E}[e^{t\boldsymbol{w}^\top \hat{\Delta}_r^\sigma \boldsymbol{w}}] = \prod_{r=1}^{n/2} \mathbb{E}[e^{\frac{t}{n/2}\boldsymbol{w}^\top (S_r^\sigma S_r^{\sigma T} - T)\boldsymbol{w}}] \leq e^{\frac{2t^2 M^2 \pi}{n}}.$$

Then by Jensen's inequality,

$$\mathbb{E}[e^{t\boldsymbol{w}^\top \Delta \boldsymbol{w}}] = \mathbb{E}[e^{t\boldsymbol{w}^\top \frac{1}{|S_n|}\sum_{\sigma \in S_n} \sum_{r=1}^{n/2} \hat{\Delta}_r^\sigma \boldsymbol{w}}] \leq \frac{1}{|S_n|} \sum_{\sigma \in S_n} \mathbb{E}[e^{t\boldsymbol{w}^\top \sum_{r=1}^{n/2} \hat{\Delta}_r^\sigma \boldsymbol{w}}] \leq e^{\frac{2t^2 M^2 \pi}{n}}.$$

Therefore, by the property of sub-gaussian random variable, for any $\boldsymbol{w} \in S^{d-1}$,

$$P(\boldsymbol{w}^\top \Delta \boldsymbol{w} > t) \le e^{-\frac{nt^2}{2M^2\pi}}.$$

Then by (34) and let $\epsilon = 1/2$, we have

$$P(||\Delta||_2 > t) \le 3^d e^{-\frac{nt^2}{2M^2\pi}} = e^{d\log 3 - \frac{nt^2}{2M^2\pi}}.$$

Let $t = \sqrt{(2\pi \log 3M^2)\frac{d+t}{n}}$, then with probability at least $1 - e^{-t}$,

$$||\Delta|| \lesssim \sqrt{\frac{d+t}{n}}.$$

3. By (2), for any $A \subset [p]$ with $|A| = s$, with probability at least $1 - e^{-t}$,

$$\frac{||\Delta_{A\times A}||}{||\Sigma_{A\times A}||} \lesssim \sqrt{\frac{s+t}{n}}.$$

Therefore

$$\frac{||\Delta||_{2,s}}{||\Sigma||_{2,s}} = \frac{\sup_{|A|=s} ||\Delta_{A\times A}||}{\sup_{|A|=s} ||\Sigma_{A\times A}||} \le \sup_{|A|=s} \frac{||\Delta_{A\times A}||}{||\Sigma_{A\times A}||} \lesssim \sqrt{\frac{s+t}{n}}$$

with probability at least $1 - \binom{p}{s} e^{-t} = 1 - e^{s\log p - t}$.

This implies with probability at least $1 - e^{-t}$,

$$\frac{||\Delta||_{2,s}}{||\Sigma||_{2,s}} \lesssim \sqrt{\frac{s\log p + t}{n}} \Rightarrow ||\Delta||_{2,s} \lesssim ||\Sigma||_{2,s}\sqrt{\frac{s\log p + t}{n}}.$$

□

## Proof of Lemma 6.3

Recall that $\sigma^2_{g_1(\boldsymbol{u}_i)} = \mathrm{Var}(g_1(Z; \boldsymbol{u}_i)) = x(\boldsymbol{u}_i)^\top \Sigma_{h_Z} x(\boldsymbol{u}_i)$ and for $i = 1, 2, ..., p$, $H_i = \boldsymbol{u}_i^\top \hat{\Sigma} \boldsymbol{v}_0$. Taylor expansion yields

$$\hat{\Sigma} - \Sigma = \sin(\frac{\pi}{2}\hat{T}) - \sin(\frac{\pi}{2}T) = \cos(\frac{\pi}{2}T) \circ (\hat{T} - T) \cdot \frac{\pi}{2} - \frac{1}{2}\sin(\frac{\pi}{2}\check{T}) \circ \frac{\pi}{2}(\hat{T} - T) \circ \frac{\pi}{2}(\hat{T} - T).$$

It follows

$$\begin{aligned} H_i - \mathbb{E}[H_i] =& \boldsymbol{u}_i^\top (\hat{\Sigma} - \Sigma)\boldsymbol{v}_0 \\ =& \frac{\pi}{2}\boldsymbol{u}_i^\top \cos(\frac{\pi}{2}T) \circ (\hat{T} - T)\boldsymbol{v}_0 - \frac{1}{2}\boldsymbol{u}_i^\top \sin(\frac{\pi}{2}\check{T}) \circ \frac{\pi}{2}(\hat{T} - T) \circ \frac{\pi}{2}(\hat{T} - T)\boldsymbol{v}_0. \end{aligned} \tag{35}$$

Let $J_i = \boldsymbol{u}_i^\top \cos(\frac{\pi}{2}T) \circ \hat{T}\boldsymbol{v}_0 = \frac{1}{n(n-1)/2}\sum_{i<i'} g(\boldsymbol{Z}_i, \boldsymbol{Z}_{i'}; \boldsymbol{u}_i)$, which is a U-statistics of order two. Then by the Berry-Essen bound for U statistics [6], we have

$$\sup_{x\in\mathbb{R}} |P(\frac{\sqrt{n}(J_i - \mathbb{E}[J_i])}{2\sigma_{g_1(\boldsymbol{u}_i)}} \leq x) - \Phi(x)| \leq C\frac{\eta_g^3}{\sigma_{g_1(\boldsymbol{u}_i)}^3} \cdot \frac{1}{\sqrt{n}}, \tag{36}$$

where $\eta_g^3 = \mathbb{E}[|g(\boldsymbol{Z}, \boldsymbol{Z}'; \boldsymbol{u}_i)^3|]$.

To prove Lemma 6.3, it is sufficient to show that $\frac{\eta_g^3}{\sigma_{g_1}^3}$ is upper bounded by a constant. If we can show this, then apply Lemma 6.2 and (36) to (35), we would get the desired result.

Therefore we proceed to prove that $\frac{\eta_g^3}{\sigma_{g_1}^3}$ is upper bounded by a constant. Recall that

$$\sigma_{g_1}^2 = \mathrm{Var}(g_1(\boldsymbol{Z}; \boldsymbol{u}_i)) = \mathrm{Var}(\mathbb{E}[g(\boldsymbol{Z}, \boldsymbol{Z}'; \boldsymbol{u}_i)|\boldsymbol{Z}]) = \mathrm{Var}(g(\boldsymbol{Z}, \boldsymbol{Z}')) - E[\mathrm{Var}(g(\boldsymbol{Z}, \boldsymbol{Z}')|\boldsymbol{Z})],$$

$h_Z(\boldsymbol{Z}) = \mathbb{E}[h(\boldsymbol{Z}, \boldsymbol{Z}')|\boldsymbol{Z}]$, and $\Sigma_{h_Z} = \mathrm{Var}(h_Z(\boldsymbol{Z}))$ for $h_Z(\boldsymbol{Z}) = \mathbb{E}[\mathrm{sgn}(\boldsymbol{Z} - \boldsymbol{Z}') \otimes \mathrm{sgn}(\boldsymbol{Z} - \boldsymbol{Z}')|\boldsymbol{Z}] = \mathbb{E}[\mathrm{vec}(\hat{T}_{\boldsymbol{Z},\boldsymbol{Z}'})|\boldsymbol{Z}]$, where $\hat{T}_{\boldsymbol{Z},\boldsymbol{Z}'}$ is the Kendall's tau estimator based on only two samples $\{\boldsymbol{Z}, \boldsymbol{Z}'\}$.

We start from

$$\begin{aligned}\sigma_{g_1}^2 &= \mathrm{Var}(g_1(\boldsymbol{Z}; \boldsymbol{u}_i)) = \mathrm{vec}(\boldsymbol{u}_i\boldsymbol{v}_0^\top \circ \cos(\frac{\pi}{2}T))^\top \cdot \Sigma_{h_Z} \cdot \mathrm{vec}(\boldsymbol{u}_i\boldsymbol{v}_0^\top \circ \cos(\frac{\pi}{2}T)) \\ &\geq M_3||(0, \boldsymbol{u}_i)(1, -\boldsymbol{\beta})^\top \circ \cos(\frac{\pi}{2}T)||_F^2 \geq \frac{M_3}{M_1^2}||\boldsymbol{u}_i||_2^2.\end{aligned}$$

The second inequality uses the fact that for any $i \neq j$,

$$\cos(\frac{\pi}{2}T_{ij}) = \sqrt{1 - \Sigma_{ij}^2} = \det(\Sigma_{\{i,j\},\{i,j\}}) \geq \lambda_{\min}(\Sigma_{\{i,j\},\{i,j\}}) \geq 1/M_1.$$

Therefore, we have $\sigma_{g_1}^3 \geq \frac{M_3^{3/2}}{M_1^3}||\boldsymbol{u}||_2^3$. Then we derive the upper bound for $\eta_g^3$. Taylor expansion yields $\cos(\frac{\pi}{2}T) = \sum_{k=0}^\infty \binom{1/2}{k}(-1)^k \Sigma \circ_{2k} \Sigma$. Then

$$||\cos(\frac{\pi}{2}T)||_2 \leq \sum_{k=0}^\infty |\binom{1/2}{k}| \cdot ||\Sigma \circ_{2k} \Sigma||_2 \leq \sum_{k=0}^\infty |\binom{1/2}{k}| \cdot ||\Sigma||,$$

where the last inequality comes from Theorem 5.5.18 in [11].

Therefore, $||\cos(\frac{\pi}{2}T)||_2 \leq \sum_{k=0}^\infty |\binom{1/2}{k}| \cdot ||\Sigma|| = 2||\Sigma|| \leq 2M_1$. In addition, since $\boldsymbol{\beta} = \Sigma_{XX}^{-1}\Sigma_{XY}$, we have

$$||\boldsymbol{\beta}||_2 = ||\Sigma_{XX}^{-1}\Sigma_{XY}||_2 \leq M_1||\Sigma_{XY}||_2 \leq M_1^2.$$

This implies

$$
\begin{aligned}
\eta_{g_1}^3 &= \mathbb{E}[|g(\boldsymbol{Z}, \boldsymbol{Z}'; \boldsymbol{u}_i)|^3] = \mathbb{E}[|\mathrm{sgn}(\boldsymbol{Z} - \boldsymbol{Z}')^\top (\boldsymbol{u}_i \boldsymbol{v}_0^\top \circ \cos(\frac{\pi}{2}T))\mathrm{sgn}(\boldsymbol{Z} - \boldsymbol{Z}')|^3] \\
&= \mathbb{E}[|\mathrm{tr}(\boldsymbol{u}_i \boldsymbol{v}_0^\top \mathrm{diag}(\mathrm{sgn}(\boldsymbol{Z} - \boldsymbol{Z}')) \cos(\frac{\pi}{2}T)\mathrm{diag}(\mathrm{sgn}(\boldsymbol{Z} - \boldsymbol{Z}')))|^3] \\
&= \mathbb{E}[|\mathrm{tr}(\boldsymbol{v}_0^\top \mathrm{diag}(\mathrm{sgn}(\boldsymbol{Z} - \boldsymbol{Z}')) \cos(\frac{\pi}{2}T)\mathrm{diag}(\mathrm{sgn}(\boldsymbol{X} - \boldsymbol{X}'))\boldsymbol{u}_i)|^3] \\
&= \mathbb{E}[|\boldsymbol{v}_0^\top \mathrm{diag}(\mathrm{sgn}(\boldsymbol{Z} - \boldsymbol{Z}')) \cos(\frac{\pi}{2}T)\mathrm{diag}(\mathrm{sgn}(\boldsymbol{Z} - \boldsymbol{Z}'))\boldsymbol{u}_i|^3] \\
&= \mathbb{E}[\Big| ||\boldsymbol{u}_i||_2 ||\boldsymbol{v}_0||_2 || \cos(\frac{\pi}{2}T)||_2 \Big|^3] \\
&\le \mathbb{E}[\Big| ||\boldsymbol{u}_i||_2 M_1^2 \cdot 2M_1 \Big|^3] \\
&\le 8M_1^9 ||\boldsymbol{u}_i||_2^3.
\end{aligned}
$$

The last equality due to the fact that

$$
||A||_2 = \sup_{\boldsymbol{u}, \boldsymbol{v}_0 : ||\boldsymbol{u}|| = ||\boldsymbol{v}_0|| = 1} \boldsymbol{u}^\top A \boldsymbol{v}_0 = \sup_{\boldsymbol{u}, \boldsymbol{v}_0 : ||\boldsymbol{u}|| = ||\boldsymbol{v}_0|| = 1} (\boldsymbol{u}^\top D) A (D\boldsymbol{v}_0) = ||DAD||_2,
$$

when diagonal elements of $D$ is either 1 or $-1$.

It follows

$$
\frac{\eta_g^3}{\sigma_{g_1}^3} \le \frac{8M_1^9 ||\boldsymbol{u}_i||_2^3}{\frac{M_3^{3/2}}{M_1^3} ||\boldsymbol{u}_i||_2^3} = \frac{8M_1^{12}}{M_3^{3/2}}.
$$

Then with (36) together,

$$
\lim_{n \to \infty} \sup_{x \in \mathbb{R}} |P(\frac{\sqrt{n}(H_i - \mathbb{E}[H_i])}{2v_{g_1}} \le x) - \Phi(x)| = 0.
$$

$\square$

## Proof of Lemma 6.4

Since the $1/M_1 \le \lambda_{\min}(\Sigma) \le \lambda_{\max}(\Sigma) < M_1$, $||\Sigma^{-1}|| < M_2$, the third constraint in (19):

$$
b^{-1} n^{-a} \le ||\boldsymbol{u}||_2 \le ||\boldsymbol{u}||_1 \le b n^{a/2}
$$

is feasible when $\boldsymbol{u} = \Sigma_{i,\cdot}^{-1}$ and $a, b > 0$.

Then it is sufficient to show that $(0, \Sigma_{i,\cdot}^{-1})^\top$ satisfies the constraint condition when $\mu = C\sqrt{\frac{\log p}{n}}$ with high probability.

By Lemma 6.2.3, with probability at least $1-p^{-2}$

$$||\hat{\Sigma}_{XX}\Sigma^{-1}_{\cdot,i} - e_i^{(p)}||_\infty = ||\hat{\Sigma}_{XX}\Sigma^{-1}_{\cdot,i} - \Sigma_{XX}\Sigma^{-1}_{\cdot,i}||_\infty$$
$$\leq |\hat{\Sigma}_{XX} - \Sigma_{XX}|_\infty \cdot ||\Sigma^{-1}_{\cdot,i}||_1 \lesssim \sqrt{\frac{\log p}{n}}.$$

In addition, due to the fact that $||\Sigma^{-1}_{i,\cdot}||_1 \leq ||\Sigma^{-1}||_1 \leq M_2$, and $||\Sigma^{-1}_{i,\cdot}||_2 \geq \lambda_{\min}(\Sigma^{-1}) \geq \frac{1}{M_1}$, we concludes that the constraints in the optimization problem 19 is feasible with probability at least $1-p^{-2}$. □

## Proof of Lemma 6.5

Recall that $x(\boldsymbol{u}) = \mathrm{vec}(\boldsymbol{u}\boldsymbol{v}_0^\top \circ \cos(\frac{\pi}{2}T))$, $\sigma^2_{g_1} = x(\boldsymbol{u})^\top \Sigma_{h_Z} x(\boldsymbol{u})$, and

$$\hat{\Sigma}_{h_Z} = \frac{1}{n}\sum_{i=1}^n (\hat{h}_Z(\boldsymbol{Z}_i) - \frac{1}{n}\sum_{i'=1}^n \hat{h}_Z(\boldsymbol{Z}_{i'}))(\hat{h}_Z(\boldsymbol{Z}_i) - \frac{1}{n}\sum_{i'=1}^n \hat{h}_Z(\boldsymbol{Z}_{i'}))^\top,$$

with $\hat{h}_Z(\boldsymbol{Z}_i) = \frac{1}{n-1}\sum_{i\neq i'} \mathrm{sgn}(\boldsymbol{Z}_i - \boldsymbol{Z}_{i'}) \otimes \mathrm{sgn}(\boldsymbol{Z}_i - \boldsymbol{Z}_{i'}) \in \mathbb{R}^{d^2}$.

We would like to prove with high probability

$$|x(\boldsymbol{u})^\top (\hat{\Sigma}_{h_Z} - \Sigma_{h_Z}) x(\boldsymbol{u})| \leq \sqrt{\frac{\log p}{n^{1-2a}}}.$$

By definition, for a random vector $\boldsymbol{Z} = (Z_{(1)}, ..., Z_{(d)})$ with an independent copy $\boldsymbol{Z}'$, and any $j,k \in [d]$, let's use $h_Z(\boldsymbol{Z})_{jk}$ to denote the $[(j-1)d+k]$-th coordinate of $h_Z(\boldsymbol{Z})$

$$h_Z(\boldsymbol{Z})_{jk} = \mathbb{E}[\mathrm{sgn}(Z_{(j)} - Z'_{(j)})\mathrm{sgn}(Z_{(k)} - Z'_{(k)})|\boldsymbol{Z}],$$

and

$$\hat{h}_Z(\boldsymbol{Z}_i)_{jk} = \frac{1}{n-1}\sum_{i'\neq i}^n \mathrm{sgn}(Z_{ij} - Z_{i'j})\mathrm{sgn}(Z_{ik} - Z_{i'k}).$$

This implies $\frac{1}{n}\sum_{i=1}^n \hat{h}_Z(\boldsymbol{Z}_i)_{jk} = \hat{\tau}_{jk}$. Therefore

$$\begin{aligned}
\hat{\Sigma}_{h_Z(jk,\, j_1k_1)} =& \frac{1}{n}\sum_i (\hat{h}_Z(\boldsymbol{Z}_i)_{jk} - \frac{1}{n}\sum_{i'} \hat{h}_Z(\boldsymbol{Z}_{i'})_{jk})(\hat{h}_Z(\boldsymbol{Z}_i)_{j_1,k_1} - \frac{1}{n}\sum_{i'} \hat{h}_Z(\boldsymbol{Z}_{i'})_{j_1,k_1}) \\
=& \frac{1}{n}\sum_i (\hat{h}_Z(\boldsymbol{Z}_i)_{jk} - \hat{\tau}_{jk})(\hat{h}_Z(\boldsymbol{Z}_i)_{j_1,k_1} - \hat{\tau}_{j_1k_1}) \\
=& \frac{1}{n}\sum_i \hat{h}_Z(\boldsymbol{Z}_i)_{jk}\hat{h}_Z(\boldsymbol{Z}_i)_{j_1,k_1} - \hat{\tau}_{jk}\hat{\tau}_{j_1k_1}.
\end{aligned}$$

It follows

$$
\begin{aligned}
x(\boldsymbol{u})^\top \hat{\Sigma}_{h_Z} x(\boldsymbol{u}) =& \frac{1}{n}\sum_i x(\boldsymbol{u})^\top \Big(\hat{h}_Z(\boldsymbol{Z}_i) - \frac{1}{n}\sum_{i'} \hat{h}_Z(\boldsymbol{Z}_{i'})\Big)\Big(\hat{h}_Z(\boldsymbol{Z}_i) - \frac{1}{n}\sum_{i'} \hat{h}_Z(\boldsymbol{Z}_{i'})\Big)^\top x(\boldsymbol{u}) \\
=& \frac{1}{n}\sum_i x(\boldsymbol{u})^\top \hat{h}_Z(\boldsymbol{Z}_i)\hat{h}_Z(\boldsymbol{Z}_i)^\top x(\boldsymbol{u}) - x(\boldsymbol{u})^\top \mathrm{vec}(\hat{T})\mathrm{vec}(\hat{T})^\top x(\boldsymbol{u}) \\
=& \frac{1}{n}\sum_i [x(\boldsymbol{u})^\top \hat{h}_Z(\boldsymbol{Z}_i)]^2 - [x(\boldsymbol{u})^\top \mathrm{vec}(\hat{T})]^2.
\end{aligned}
$$

Since

$$
x(\boldsymbol{u})^\top \hat{h}_Z(\boldsymbol{Z}_i) = \frac{1}{n-1}\sum_{i\neq i'} x(\boldsymbol{u})^\top \Big(\mathrm{sgn}(\boldsymbol{Z}_i - \boldsymbol{Z}_{i'}) \otimes \mathrm{sgn}(\boldsymbol{Z}_i - \boldsymbol{Z}_{i'})\Big).
$$

Conditional on $\boldsymbol{Z}_i$, $x(\boldsymbol{u})^\top \mathrm{sgn}(\boldsymbol{Z}_i - \boldsymbol{Z}_{i'}) \otimes \mathrm{sgn}(\boldsymbol{Z}_i - \boldsymbol{Z}_{i'})$ are $n-1$ *i.i.d* random vectors. In addition, similar as the proof in Lemma 6.3,

$$
\begin{aligned}
|x(\boldsymbol{u})^\top \mathrm{sgn}(\boldsymbol{Z}_i - \boldsymbol{Z}_{i'}) \otimes \mathrm{sgn}(\boldsymbol{Z}_i - \boldsymbol{Z}_{i'})| = & |\mathrm{sgn}(\boldsymbol{Z} - \boldsymbol{Z}')^\top (\boldsymbol{u}\boldsymbol{v}_0^\top \circ \cos(\frac{\pi}{2}T))\mathrm{sgn}(\boldsymbol{Z} - \boldsymbol{Z}')| \\
=& |\mathrm{tr}(\boldsymbol{u}\boldsymbol{v}_0^\top \mathrm{diag}(\mathrm{sgn}(\boldsymbol{Z} - \boldsymbol{Z}')) \cos(\frac{\pi}{2}T) diag(\mathrm{sgn}(\boldsymbol{Z} - \boldsymbol{Z}')))| \\
=& |\boldsymbol{v}_0^\top \mathrm{diag}(\mathrm{sgn}(\boldsymbol{Z} - \boldsymbol{Z}')) \cos(\frac{\pi}{2}T)\mathrm{diag}(\mathrm{sgn}(\boldsymbol{Z} - \boldsymbol{Z}'))\boldsymbol{u}| \\
\leq& ||\boldsymbol{v}_0||_2 || \cos(\frac{\pi}{2}T)||_2 ||\boldsymbol{u}||_2 \leq 2M_1^3 ||\boldsymbol{u}||_2.
\end{aligned}
$$

Therefore, by Hoeffding inequality,

$$
P(|x(\boldsymbol{u})^\top \hat{h}_Z(\boldsymbol{Z}_i) - x(\boldsymbol{u})^\top h_Z(\boldsymbol{Z}_i)| > t | Z_i) \leq e^{-\frac{nt^2}{4M_1^3 ||\boldsymbol{u}||_2}}.
$$

This implies

$$
\mathbb{E}[e^{t(x(\boldsymbol{u})^\top \hat{h}_Z(\boldsymbol{Z}_i) - x(\boldsymbol{u})^\top h_Z(\boldsymbol{Z}_i))}] \leq e^{\frac{4M_1^3 ||\boldsymbol{u}||_2 t^2}{n}}.
$$

Therefore the sub-gaussian norm of $x(\boldsymbol{u})^\top \hat{h}_Z(\boldsymbol{Z}_i) - x(\boldsymbol{u})^\top h_Z(\boldsymbol{Z}_i)$,

$$
||x(\boldsymbol{u})^\top \hat{h}_Z(\boldsymbol{Z}_i)) - x(\boldsymbol{u})^\top h_Z(\boldsymbol{Z}_i)||_{\psi_2} \leq \frac{4M_1^3 ||\boldsymbol{u}||_2}{n},
$$

and this implies

$$
\begin{aligned}
& ||\frac{1}{n}\sum_{i=1}^n x(\boldsymbol{u})^\top \hat{h}_Z(\boldsymbol{Z}_i) - x(\boldsymbol{u})^\top h_Z(\boldsymbol{Z}_i)||_{\psi_2} \\
\leq& \frac{1}{n}\sum_{i=1}^n ||x(\boldsymbol{u})^\top \hat{h}_Z(\boldsymbol{Z}_i) - x(\boldsymbol{u})^\top h_Z(\boldsymbol{Z}_i)||_{\psi_2} \leq \frac{4M_1^3 ||\boldsymbol{u}||_2}{n}.
\end{aligned}
$$

Therefore

$$P(|\frac{1}{n}\sum_{i=1}^{n} x(\boldsymbol{u})^\top \hat{h}_Z(\boldsymbol{Z}_i) - x(\boldsymbol{u})^\top h_Z(\boldsymbol{Z}_i)| > t) \leq e^{-\frac{nt^2}{4M^3||\boldsymbol{u}||_2}}.$$

Similarly, by Hoeffding inequality,

$$\begin{aligned}
P(|x(\boldsymbol{u})^\top \text{vec}(\hat{T}) - x(\boldsymbol{u})^\top \text{vec}(T)| > t) \leq & e^{-\frac{nt^2}{4M_1^3||\boldsymbol{u}||_2}}, \\
P(|\frac{1}{n}\sum_{i=1}^{n}(x(\boldsymbol{u})^\top h_Z(\boldsymbol{Z}_i))^2 - \mathbb{E}[(x(\boldsymbol{u})^\top h_Z(\boldsymbol{Z}_i))^2]| > t) \leq & e^{-\frac{nt^2}{8M_1^6||\boldsymbol{u}||_2^2}}.
\end{aligned}$$

It follows with probability at least $1 - 4e^{-\frac{nt^2}{4M_1^3||\boldsymbol{u}||_2}} - e^{-\frac{nt^2}{8M_1^6||\boldsymbol{u}||_2^2}}$,

$$\begin{aligned}
& |x(\boldsymbol{u})^\top (\hat{\Sigma}_{h_Z} - \Sigma_{h_Z}) x(\boldsymbol{u})| \\
=& |\frac{1}{n}\sum_i [x(\boldsymbol{u})^\top \hat{h}_Z(\boldsymbol{Z}_i)]^2 - [x(\boldsymbol{u})^\top \text{vec}(\hat{T})]^2 - \frac{1}{n}\sum_i \mathbb{E}[(x(\boldsymbol{u})^\top h_Z(\boldsymbol{Z}_i))^2] + [x(\boldsymbol{u})^\top \text{vec}(T)]^2 \\
=& |\frac{1}{n}\sum_i [x(\boldsymbol{u})^\top \hat{h}_Z(\boldsymbol{Z}_i)]^2 - [x(\boldsymbol{u})^\top \text{vec}(\hat{T})]^2 - \frac{1}{n}\sum_i (x(\boldsymbol{u})^\top h_Z(\boldsymbol{Z}_i))^2 + \frac{1}{n}\sum_i (x(\boldsymbol{u})^\top h_Z(\boldsymbol{Z}_i))^2 \\
& - \frac{1}{n}\sum_i \mathbb{E}[(x(\boldsymbol{u})^\top h_Z(\boldsymbol{Z}_i))^2] + [x(\boldsymbol{u})^\top \text{vec}(T)]^2| \\
\leq& |\frac{1}{n}\sum_i (x(\boldsymbol{u})^\top \hat{h}_Z(\boldsymbol{Z}_i))^2 - \frac{1}{n}\sum_i (x(\boldsymbol{u})^\top h_Z(\boldsymbol{Z}_i))^2| + |\frac{1}{n}\sum_i (x(\boldsymbol{u})^\top h_Z(\boldsymbol{Z}_i))^2 - \frac{1}{n}\sum_i \mathbb{E}[(x(\boldsymbol{u})^\top h_Z(\boldsymbol{Z}_i))^2]| \\
& + |[x(\boldsymbol{u})^\top \text{vec}(\hat{T})]^2 - [x(\boldsymbol{u})^\top \text{vec}(T)]^2| \\
=& |\frac{1}{n}\sum_i (x(\boldsymbol{u})^\top \hat{h}_Z(\boldsymbol{Z}_i) - x(\boldsymbol{u})^\top h_Z(\boldsymbol{Z}_i))(x(\boldsymbol{u})^\top \hat{h}_Z(\boldsymbol{Z}_i) + x(\boldsymbol{u})^\top h_Z(\boldsymbol{Z}_i))| + |[x(\boldsymbol{u})^\top \text{vec}(\hat{T})]^2 - [x(\boldsymbol{u})^\top \text{vec}(T)]^2| \\
& + |\frac{1}{n}\sum_i (x(\boldsymbol{u})^\top h_Z(\boldsymbol{Z}_i) - \mathbb{E}[(x(\boldsymbol{u})^\top h_Z(\boldsymbol{Z}_i)])((x(\boldsymbol{u})^\top h_Z(\boldsymbol{Z}_i) + \mathbb{E}[(x(\boldsymbol{u})^\top h_Z(\boldsymbol{Z}_i)])|.
\end{aligned}$$

In addition,

$$|x(\boldsymbol{u})^\top h_Z(\boldsymbol{Z}_i)| \leq ||\text{vec}(\boldsymbol{u}\boldsymbol{v}_0^\top \circ \cos(\frac{\pi}{2}T))||_1 \leq ||\boldsymbol{u}||_1 ||\boldsymbol{v}_0||_1 \leq ||\boldsymbol{u}||_1 ||\boldsymbol{\beta}||_1 \lesssim \sqrt{s}||\boldsymbol{u}||_1.$$

Similarly,

$$|x(\boldsymbol{u})^\top \hat{h}_Z(\boldsymbol{Z}_i)| \lesssim \sqrt{s}||\boldsymbol{u}||_1, |x(\boldsymbol{u})^\top \text{vec}(\hat{T})| \lesssim \sqrt{s}||\boldsymbol{u}||_1, \text{ and } |x(\boldsymbol{u})^\top \text{vec}(T)| \lesssim \sqrt{s}||\boldsymbol{u}||_1.$$

Therefore

$$\begin{aligned}
&|x(\boldsymbol{u})^\top(\hat{\Sigma}_{h_Z} - \Sigma_{h_Z})x(\boldsymbol{u})| \\
\leq&\sqrt{s}||\boldsymbol{u}||_1 \cdot \Big[|\frac{1}{n}\sum_i x(\boldsymbol{u})^\top \hat{h}_Z(\boldsymbol{Z}_i) - x(\boldsymbol{u})^\top h_Z(\boldsymbol{Z}_i)| + |x(\boldsymbol{u})^\top \mathrm{vec}(\hat{T}) - x(\boldsymbol{u})^\top \mathrm{vec}(T)| \\
&+ |\frac{1}{n}\sum_i x(\boldsymbol{u})^\top h_Z(\boldsymbol{Z}_i) - \mathbb{E}[(x(\boldsymbol{u})^\top h_Z(\boldsymbol{Z}_i)]|\Big] \\
\leq&3\sqrt{s}||\boldsymbol{u}||_1 t.
\end{aligned}$$

Therefore,

$$P(|x(\boldsymbol{u})^\top(\hat{\Sigma}_{h_Z} - \Sigma_{h_Z})x(\boldsymbol{u})| \gtrsim \sqrt{s}||\boldsymbol{u}||_1 t) \leq 4e^{-\frac{nt^2}{4M_1^3||\boldsymbol{u}||_2}} + e^{-\frac{nt^2}{8M_1^6||\boldsymbol{u}||_2^2}}.$$

Let $t = \sqrt{\frac{8M_1^3 \log p \cdot n^a}{n}}$, and by the fact that $||\boldsymbol{u}||_2 \leq ||\boldsymbol{u}||_1 \leq n^{a/2}$, we have for any $\epsilon > 0$,

$$P(|x(\boldsymbol{u})^\top(\hat{\Sigma}_{h_Z} - \Sigma_{h_Z})x(\boldsymbol{u})| \gtrsim \sqrt{\frac{s \log p}{n^{1-2a}}}) \leq 5p^{-2}.$$

□

## Proof of Lemma 6.6

Recall that $x(\boldsymbol{u}) = \mathrm{vec}(\boldsymbol{u}\boldsymbol{v}_0^\top \circ \cos(\frac{\pi}{2}T))$, and $\hat{x}(\boldsymbol{u}) = \mathrm{vec}(\boldsymbol{u}\hat{\boldsymbol{v}_0}^\top \circ \cos(\frac{\pi}{2}\hat{T}))$, therefore

$$\begin{aligned}
||x(\boldsymbol{u}) - \hat{x}(\boldsymbol{u})||_1 =&||\mathrm{vec}(\boldsymbol{u}\boldsymbol{v}_0^\top \circ \cos(\frac{\pi}{2}T)) - \mathrm{vec}(\boldsymbol{u}\hat{\boldsymbol{v}}^\top \circ \cos(\frac{\pi}{2}\hat{T}))||_1 \\
=&|\boldsymbol{u}\boldsymbol{v}_0^\top \circ \cos(\frac{\pi}{2}T) - \boldsymbol{u}\hat{\boldsymbol{v}}^\top \circ \cos(\frac{\pi}{2}\hat{T})||_1 \\
\leq&||\boldsymbol{u}(\boldsymbol{v}_0 - \hat{\boldsymbol{v}})^\top||_1 \leq ||\boldsymbol{u}||_1||\boldsymbol{v}_0 - \hat{\boldsymbol{v}}||_\infty \\
\leq&||\boldsymbol{u}||_1||\boldsymbol{v}_0 - \hat{\boldsymbol{v}}||_2 \lesssim n^a\sqrt{\frac{s \log p}{n}}.
\end{aligned}$$

□

## Proof of Lemma 6.7

Since $||\boldsymbol{\beta}||_2 = ||\Sigma_{XX}^{-1}\Sigma_{XY}||_2 \geq M_1^{-2}$, then

$$\begin{aligned}
\sigma_{g_1(\boldsymbol{u})}^2 &= \mathrm{Var}(g_1(\boldsymbol{Z};\boldsymbol{u})) = \mathrm{vec}(\boldsymbol{u}\boldsymbol{v}_0^\top \circ \cos(\frac{\pi}{2}T))^\top \cdot \Sigma_{h_Z} \cdot \mathrm{vec}(\boldsymbol{u}\boldsymbol{v}_0^\top \circ \cos(\frac{\pi}{2}T)) \\
&\geq \frac{M_3}{M_1}||(0, \boldsymbol{u}[2:p])(1, -\boldsymbol{\beta})^\top \circ \cos(\frac{\pi}{2}T)||_F^2 \\
&\geq \frac{M_3}{M_1^5}||\boldsymbol{u}||_2^2 \geq \frac{M_3}{M_1^5 n^{2a}}.
\end{aligned}$$

□